\documentclass{aa}
\usepackage{graphicx,natbib}
\usepackage{txfonts}
\usepackage{colordvi}
\begin{document}
   \title{Testing the inverse-Compton catastrophe scenario in the intra-day variable blazar S5\,0716+71} 
    
   \subtitle{III. Rapid and correlated flux density variability from radio to sub-mm bands}

     \author{L. Fuhrmann\inst{1,2,3}, T. P. Krichbaum\inst{1}, A. Witzel\inst{1}, A. Kraus\inst{1}, 
       S. Britzen\inst{1}, S. Bernhart\inst{1}, C. M. V. Impellizzeri\inst{1}, I. Agudo\inst{1,4}, J. Klare\inst{1}, 
       B. W. Sohn\inst{1,5}, E. Angelakis\inst{1}, U. Bach\inst{1,3}, K. \'E. Gab\'anyi\inst{1,6,7}, 
       E. K\"ording\inst{8}, A. Pagels\inst{1}, J. A. Zensus\inst{1}, S. J. Wagner\inst{9}, L. Ostorero\inst{10,11}, 
       H. Ungerechts\inst{12}, M. Grewing\inst{13}, M. Tornikoski\inst{14}, A. J. Apponi\inst{15}, 
       B. Vila-Vilar\'o\inst{16}, L. M. Ziurys\inst{15}, R. G. Strom\inst{17}
} 
   \offprints{L. Fuhrmann,\\ e-mail: lfuhrmann@mpifr-bonn.mpg.de}

   \institute{Max-Planck-Institut f\"ur Radioastronomie, Auf dem H\"ugel 69, 53121 Bonn, Germany
              \and
             Dipartimento di Fisica, Universit\`a di Perugia, Via A. Pascoli, 06123 Perugia, Italy
             \and
             INAF, Osservatorio Astronomico di Torino, via Osservatorio 20, 10025 Pino Torinese (TO), Italy
             \and 
	     Instituto de Astrof\'{\i}sica de Andaluc\'{\i}a, CSIC, Apartado 3004, 18080 Granada, Spain
	     \and
	     Korea Astronomy \& Space Science Institute, 61-1 Hwaam-dong, 305-348 Daejeon
	     \and
	     Hungarian Academy of Sciences Research Group for Physical Geodesy and Geodynamics, Budapest, Hungary
	     \and
	     F\"OMI Satellite Geodetic Observatory, Budapest, Hungary
	     \and 
	     School of Physics \& Astronomy, University of Southampton, Southampton, Hampshire, SO17 1BJ, United Kingdom 
	     \and
             Landessternwarte Heidelberg-K\"onigstuhl, K\"onigstuhl, D-69117 Heidelberg, Germany 
             \and
	     Dipartimento di Fisica Generale ``Amedeo Avogadro'', Universit\`a degli Studi di Torino, Via P. Giuria 1, 10125 TORINO, Italy
	     \and
	     Istituto Nazionale di Fisica Nucleare (INFN), Sezione di Torino, Via P. Giuria 1, 10125 Torino, Italy
	     \and
             Institut de Radio Astronomie Millim\'etrique, Avenida Divina Pastora 7, Local 20, 18012 Granada, Spain
	     \and
             Institut de Radio Astronomie Millim\'etrique, 300 Rue de la Piscine, Domaine Universitaire de Grenoble, 
	     St. Martin d'H\`eres, F-38406, France  
	     \and
             Mets\"ahovi Radio Observatory, Helsinki University of Technology, Mets\"ahovintie 114, 02540 Kylm\"al\"a, Finland
             \and
             Arizona Radio Observatory, University of Arizona, 933 N. Cherry Avenue, Tucson, AZ 85721 USA
	     \and
             University of Arizona, Steward Observatory, 933 N. Cherry Ave., Tucson, AZ 85721, USA 
             \and
             ASTRON, Postbus 2, 7990 AA Dwingeloo; and Astronomical Institute, University of Amsterdam, The Netherlands     
}
   \date{the date should be inserted later}
   \authorrunning{L. Fuhrmann et al.}
 \abstract
{}
{The BL Lac object S5\,0716+71 was observed in a global multi-frequency campaign    
to search for rapid and correlated flux density variability and signatures of an 
inverse-Compton (IC) catastrophe during the states of extreme apparent brightness 
temperatures.} 
{The observing campaign involved simultaneous ground-based monitoring at radio to 
IR/optical wavelengths and was centered around a 500-ks pointing with the INTEGRAL 
satellite (November 10--17, 2003). Here, we present the combined analysis and results 
of the radio observations, covering the cm- to sub-mm bands. This facilitates a detailed 
study of the variability characteristics of an inter- to intra-day variable IDV source 
from cm- to the short mm-bands. We further aim to constrain the variability brightness 
temperatures ($T_{B}$) and Doppler factors ($\delta$) comparing the radio-bands with the hard
X-ray emission, as seen by INTEGRAL at 3--200\,keV.}
{0716+714 was in an exceptionally high state and different (slower) phase of short-term variability, 
when compared to the past, most likely due to a pronounced outburst shortly before the campaign. 
The flux density variability in the cm- to mm-bands is dominated by a $\sim 4$\,day time 
scale amplitude increase of up to $\sim\,35$\,\%, systematically more pronounced towards 
shorter wavelengths. The cross-correlation analysis reveals systematic time-lags with the 
higher frequencies varying earlier, similar to canonical variability on longer time-scales. 
The increase of the variability amplitudes with frequency contradicts expectations from 
standard interstellar scintillation (ISS) and suggests a source-intrinsic origin for the 
observed inter-day variability. We find an inverted synchrotron spectrum peaking near 90\,GHz, 
with the peak flux increasing during the first 4 days. The lower limits to $T_{B}$ derived 
from the inter-day variations exceed the $10^{12}$\,K IC-limit by up to 3--4 orders of magnitude. 
Assuming relativistic boosting, our different estimates of $\delta$ yield robust and self-consistent 
lower limits of $\delta \geq 5 - 33$ -- in good agreement with $\delta_{VLBI}$ obtained from VLBI studies 
and the IC-Doppler factors $\delta_{IC}\,>$\,14--16 obtained from the INTEGRAL data.} 
{The non-detection of S5\,0716+714 with INTEGRAL in this campaign excludes an excessively high 
X-ray flux associated with a simultaneous IC catastrophe. Since a strong contribution from 
ISS can be excluded, we conclude that relativistic Doppler boosting naturally explains the 
apparent violation of the theoretical limits. All derived Doppler factors are internally 
consistent, agree with the results from different observations and can be explained within 
the framework of standard synchrotron-self-Compton (SSC) jet models of AGN.}

   \maketitle
%
  
\section{Introduction}\label{intro}
Rapid intensity and polarization variability on time scales of hours to days 
is frequently observed in compact blazar cores at centimeter wavelengths. Since its 
discovery in 1985 such IntraDay variability (IDV, Witzel et al. 1986, Heeschen et al. 
1987) has been found in a significant fraction ($\sim$\,20--30\%) of flat-spectrum 
quasars and BL\,Lac objects (Quirrenbach et al. 1992, Kedziora-Chudczer et al. 2001, 
Lovell et al. 2003). The `classical' IDV sources \citep[of type II,][]{1987AJ.....94.1493H} 
show flux density variations in the radio bands on time scales of $\lesssim$\,0.5\,--2\,days 
and variability amplitudes ranging from a few up to 30\,\%. Stronger and often faster variability 
is seen in both linear polarization \citep[e.g.][]{1989A&A...226L...1Q,2003A&A...401..161K} 
and more recently also in circular polarization \citep[][]{2000ApJ...538..623M}.

When the observed IDV is interpreted as being source-intrinsic, the apparent variability brightness 
temperatures of $T_{B}\,\geq\,10^{18}$\,K, largely exceed the inverse Compton (IC) limit of 
$10^{12}$\,K \citep[][]{1969ApJ...155L..71K,1994ApJ...426...51R} by several orders of magnitude 
\citep[see also][]{1995ARA&A..33..163W}. Consequently, extreme relativistic Doppler-boosting
factors ($\delta\,>$\,100) would be required to bring these values down to the IC-limit.
As such high Doppler-factors (and related bulk Lorentz factors) are not observed in AGN jets 
with VLBI, alternate jet (and shock-in-jet) models use non-spherical geometries, which allow 
for additional relativistic corrections \citep[e.g.][]{1999ApJ...511..136S,1991A&A...241...15Q}. 
It is also possible that the brightness temperatures are intrinsically high and coherent emission 
processes are involved \citep[][]{1992ApJ...391L..59B,1998MNRAS.301..414B}. Another possibility 
to explain intrinsic brightness temperatures of $>\,10^{12}$\,K is a more homogeneous ordering of the 
magnetic field \citep[][]{2006ChJAA...6..530Q}.

Owing to the small source sizes in compact flat-spectrum AGN, interstellar scintillation 
(ISS) is an unavoidable process. For a small number of more rapid, intra-hour variable sources, 
seasonal cycles and/or variability pattern arrival time delays are observed. This
is used as convincing argument that such rapid variability is caused by ISS
\citep[PKS\,0405$-$385, PKS\,1257$-$326 and J1819+3845, e.g.][]
{2000aprs.conf..147J,2003ApJ...585..653B,2002Natur.415...57D}. The situation for the 
slower and `classical' (type II) IDV sources is less clear. The much longer and complex 
variability time scales and the existence of more than one characteristic variability 
time-scale, did not yet allow to unambiguously establish seasonal cycles nor arrival time 
measurements and by one of this proof that IDV of type II is solely due to ISS. 
In fact, many of these sources show an overall more complex variability behavior,
\citep[e.g. QSO 0917+624, BL 0954+658, J\,1128+5925;][]{2001ApJ...550L..11R,2001A&A...370L...9J,
2002PASA...19...64F,2004PhDT........38K,2006astro.ph.10795B,2007A&A...470...83G},
which points towards a more complicated and time-variable blending between
propagation induced source extrinsic effects and source intrinsic variability. 

In order to shed more light on the physical origin of IDV in such type II sources,
we performed a coordinated broad-band variability campaign on one of the most prominent 
IDV sources. Since the late 1980's, the compact blazar S5\,0716+71 (hereafter 0716+714) showed 
strong and fast IDV of type II whenever it was observed. In the optical band, 0716+714 is identified 
as a BL\,Lac, but has unknown redshift. A redshift of $z\,>$\,0.3 was deduced using the limits on 
the surface brightness of the host galaxy 
\citep[][]{1991ApJ...372L..71Q,1993A&AS...97..483S,1996AJ....111.2187W,2005ApJ...635..173S,2008arXiv0807.0203N}.
In this paper we will use $z\,\geq$\,0.3, which is within the measurement uncertainty of the previous 
redshift estimates. Consequently, only lower limits to any linear size and brightness temperature 
measurement can be obtained. The observed IDV time scales imply -- via the light travel time 
argument -- source sizes of only light-hours (corresponding to micro-arcsecond scales) leading to 
lower limits of $T_{B} \geq 10^{18}$\,K. Direct measurement of the nucleus with space-VLBI at 5 GHz
gives a robust lower limit of $T_{B} \ge\!2\times 10^{12}$\,K, implying a minimum equipartition Doppler 
factor of $\delta \!\ge\!4$ \citep[][]{2006A&A...452...83B}. The VLBI kinematics leads to estimates 
of the bulk jet Lorentz factor $\gamma_{\rm min}\,\geq$\,16--21, based on the measured jet speeds 
\citep[][]{2001ApJS..134..181J,2005A&A...433..815B}.  This is is unusually high for a BL\,Lac object.

0716+714 is one of the best studied blazars in the sky \citep[e.g.][]
{1996AJ....111.2187W,1998A&A...333..445G,2000A&AS..141..221Q,2003A&A...401..161K,
2005A&A...433..815B,2006A&A...452...83B}. It has been observed repeatedly in various 
multi-frequency campaigns \citep[e.g.][]{1990A&A...235L...1W,1996AJ....111.2187W,1999hxra.conf..170O,
1999A&A...351...59G,2003A&A...400..477T,2003A&A...402..151R} and is well known to be extremely variable 
($\lesssim$\,hours to months) at radio to X-ray bands 
\citep[e.g.][and references therein]{1996AJ....111.2187W,2003A&A...402..151R}. 0716+714 is so far 
the only source in which a correlation between the radio/optical variability was observed 
\citep[see][]{1991ApJ...372L..71Q,1995ARA&A..33..163W} indicating that the radio/optical emission  
has a common and source-intrinsic origin, during the observations in 1990 
\citep[][]{1996ChA&A..20...15Q,2002ChJAA...2..325Q}. Further, the detection 
of IDV at mm-wavelengths and the observed frequency dependence of the IDV 
variability amplitudes \citep[][]{2002PASA...19...14K,2003A&A...401..161K} make it  
difficult to interpret the IDV in this source solely by standard ISS.

In a source-intrinsic interpretation of IDV it is unclear if and for how long the IC-limit 
can be violated \citep[][]{1992ApJ...391..453S,2002PASA...19...77K}. The quasi-periodic and persistent 
violation of this limit (on time scales of hours to days) would imply subsequent Compton catastrophes, 
each leading to outbursts of IC scattered radiation. The efficient IC-cooling would rather quickly 
restore the local $T_{B}$ in the source, and the scattered radiation should be observable as enhanced 
emission in the X-/$\gamma$\,-ray bands \citep[][]{1969ApJ...155L..71K,1996ApJ...461..657B,2006A&A...451..797O}.
In order to search for multi-frequency signatures of such short-term IC-flashes, 0716+714 was the 
target of a global multi-frequency campaign carried out in November 2003, and centered around a 
500-ks INTEGRAL\footnote{INTErnational Gamma-Ray Astrophysics Laboratory} pointing on November 
10\,--\,17, 2003 (`core-campaign'). In order to detect contemporaneous IC-limit violations at 
lower energies, quasi-simultaneous and densely time-sampled flux density, polarization and 
VLBI monitoring observations were organized, covering the radio, millimeter, sub-millimeter, 
IR and optical\footnote{The IR/optical data were collected in collaboration with the WEBT; 
http://www.to.astro.it/blazars/webt} wavelengths.
 
Early results of this campaign including radio data at two frequencies, optical 
and INTEGRAL soft $\gamma$-ray data were presented by \citet{2006A&A...451..797O}, who also 
showed the simultaneous spectral energy distribution. The results from the mm-observations 
(3\,mm, 1.3\,mm) performed with the IRAM 30\,m radio-telescope were presented by 
\citet{2006A&A...456..117A}. The results of the VLBI observations will be presented 
in a forthcoming paper. In this paper III we present the intensity and polarization data obtained with 
the Effelsberg 100\,m radio telescope (5, 10.5, \& 32\,GHz) during the time of the core-campaign. 
We then combine these data with the (sub-) millimeter and optical data in order to determine 
the broad-band variability and spectral characteristics of 0716+714, with special regard to 
(i) the intra- to inter-day cm-/mm-variability, (ii) the brightness temperatures and IC-limit, and  
(iii) the Doppler factors combining the radio and high energy INTEGRAL data from this campaign.

\begin{table*}
\begin{center}
\caption{ The participating radio observatories, their 
  observing wavelengths/frequency, dates and total observing time $T_{Obs}$.} 
\begin{tabular}{l|lllll}
\hline
\hline
Radio telescope \& Institute  & Location              & $\lambda_{obs}$ [mm] & $\nu_{obs}$ [GHz] & dates & $T_{obs}$ [hrs]\\ 
\hline        
WSRT (14x25\,m), ASTRON       & Westerbork, NL        & 210, 180           & 1.4, 2.3        & Nov. 10--11 & 8   \\    
Effelsberg (100\,m), MPIfR    & Effelsberg, D         & 60, 28, 9          & 4.85, 10.45, 32 & Nov. 11--18 & 164 \\
Pico Veleta (30\,m), IRAM     & Granada, E            & 3.5, 1.3           & 86, 229         & Nov. 10--16 & 135 \\
Mets\"ahovi (14\,m), MRO      & Mets\"ahovi, Finland  & 8                  & 37              & Nov. 08--19 & 263 \\
Kitt Peak (12\,m), ARO        & Kitt Peak, AZ, USA    & 3                  & 90              & Nov. 14--18 & 91  \\
SMT/HHT  (10\,m), ARO         & Mt. Graham, AZ, USA   & 0.87               & 345             & Nov. 14--17 & 77  \\
JCMT (15\,m), JAC             & Mauna Kea, HI, USA    & 0.85, 0.45         & 345, 666        & Nov. 09--13 & 99 \\
\hline
\hline
\end{tabular}
\label{observatories}
\end{center}
\end{table*}

\section{Observations and Data Reduction}

\subsection{Participating Radio Observatories}\label{obs}
The intensive flux density monitoring of 0716+714 was carried out at the time of 
the core campaign between November 08\,--\,19, 2003 (JD 2452951.9\,--\,2452962.8). In 
total, 7 radio observatories were involved, covering a frequency range from 1.4 to 666\,GHz 
(wavelengths ranging from 21\,cm to 0.45\,mm). In Table \ref{observatories}, the 
participating observatories, their wavelength/frequency coverage and observing dates 
are summarized.

At the Effelsberg telescope, 0716+714 and a number of secondary calibrators 
were observed continuously at 5, 10.5, \& 32\,GHz, with a dense time sampling of about 
two flux density measurements per hour, source and frequency. The special IDV observing 
strategy and matched data reduction procedure \citep{2003A&A...401..161K} enabled high 
precision flux density and polarization measurements and facilitate the study 
of rapid variability on time scales from about 0.5\,hrs to 7 days. At 3.5 \& 1.3\,mm wavelengths 
with the IRAM 30\,m-telescope on Pico Veleta, a similar observing strategy was applied 
and is described by \citet{2006A&A...456..117A}. At Effelsberg polarization was recorded at 
$\lambda$\,6\,cm (4.85\,GHz) and $\lambda$\,2.8\,cm (10.45\,GHz), at the IRAM 30\,m-telescope 
at 3.5\,mm (86 GHz). The matched observing and calibration strategies at Effelsberg and Pico Veleta
now allow for the first time a detailed study (and e.g. cross-correlation analysis) of the short-term 
variability of an IDV source from centimeter to short-millimeter wavelengths.

\begin{table*}
\begin{center}
\caption{Parameters of the Effelsberg secondary focus receivers.} 
\begin{tabular}{l|c|c|c}
\hline
\hline
System                & 4.85\,GHz (6\,cm) & 10.45\,GHz (2.8\,cm) & 32\,GHz (9\,mm)\\
\hline 
Type                  & HEMT cooled & HEMT cooled & HEMT cooled\\
Number of Horns       & 2     & 4       & 3\\
Channels              & 4     & 8       & 12 total, 3 x correlation\\
System Temp. (zenith) & 27\,K & 50\,K   & 60\,K\\
Center frequency      & 4.85\,GHz & 10.45\,GHz  & 32\,GHz\\
RF-Filter             & 4.6\,-\,5.1\,GHz & 10.3\,-\,10.6\,GHz & 31\,-\,33\,GHz\\
IF-Bandwidth          & 500\,MHz & 300\,MHz & 2\,GHz\\
Polarization          & LHC \& RHC & LHC \& RHC & LHC \& RHC\\
Calibration           & noise diode & noise diode & noise diode\\
\hline
\hline
\end{tabular}
\label{receiver}
\end{center}
\end{table*}

Complementary, but less dense in time sampled flux density measurements were also
performed with the telescopes at Kitt Peak (12\,m), Mt. Graham (SMT/HHT 10m), and 
Mauna-Kea (JCMT, 15m), which covered the higher frequency bands (86 to 666\,GHz, see Table 1).
The Westerbork interferometer provided a few single measurements, which extends the 
frequency coverage to 1.4 and 2.3 GHz. The light curves obtained at 32\,GHz with 
the 100\,m telescope and at 37\,GHz with the Mets\"ahovi 14\,m-telescope have been 
already shown by \citet{2006A&A...451..797O}. The analysis of the 32\,GHz Effelsberg data 
has been improved.

\subsection{Effelsberg: Total Intensity}\label{eff_red}
The flux density measurements at the 100\,m radio telescope of the MPIfR 
were performed with three multi-horn receivers mounted in the secondary focus 
(see Table \ref{receiver} for details). The 4.85 and 10.5\,GHz systems are 
multi-feed heterodyne receivers designed as `software-beam-switching systems'.
Both circular polarizations (LCP, RCP) are fed into polarimeters providing the polarization 
information (Stokes Q and U) simultaneously. In contrast, the 32\,GHz receiver is designed 
as a correlation receiver (`hardware-beam-switch') and hence provides only total intensity 
information. 

The target source 0716+714 and all the calibrators are point-like within the telescope 
beam and sufficiently strong in the observed frequency range. This facilitates flux 
density measurements using cross-scans (azimuth/elevation direction) with the number of 
sub-scans matching the source brightness at the given frequency 
\citep[see][for details]{1987AJ.....94.1493H,1992A&A...258..279Q,2003A&A...401..161K}. 
Depending on the number of sub-scans (typically 4, 8 or 12), a single cross-scan lasted 
2--4 minutes. A duty cycle consisted of subsequent observation of the target source 
and several nearby secondary calibrators (steep-spectrum, point-like, non-IDV sources, e.g. 
0836+710 and 0951+699). Within such a duty-cycle, each source was observed consecutively at 
all three frequencies (in the order 32, 10.5 and 4.85\,GHz). 
During the 7 days observing period and our continuous 24\,hr/day time coverage,
we obtained for each source about 2 measurements per hour and frequency. Primary flux 
density and polarization angle calibrators (e.g. 3C\,286, see Table \ref{primecals}) were 
observed every few hours. Among our target source 0716+714, two additional IDV sources 
were observed: 0602+673 and 0917+624\footnote{The results obtained for these sources will 
be discussed elsewhere.}. 

The data reduction was done in the standard manner, and is described in e.g. 
\citet{2003A&A...401..161K,2004PhDT........38K}. It consists of the following steps:
(i) baseline subtraction and fitting of a Gaussian profile to each individual sub-scan; 
(ii) flagging of discrepant or otherwise bad sub-scans/scans; (iii) correcting the measured 
amplitudes for small residual pointing errors of the telescope ($<$\,5--20 arcseconds);  
(iv) sub-scan averaging over both slewing directions (the Q/U data and the $\lambda$\,9\,mm 
total power data were averaged before Gaussian fitting to enhanced the signal-to-noise ratio
and the quality of the Gaussian fits); (v) opacity correction using the $T_{sys}$-measurements 
obtained with each each cross-scan; and (vi) correction for remaining systematic gain-elevation 
and time-dependent gain effects using gain-transfer functions derived from the dense in time-sampled 
secondary and primary calibrator measurements assuming their stationarity. Finally, the measured 
antenna temperatures for each observed source were linked to the absolute flux-density scale, 
using the frequent primary calibrator measurements \citep[][]{1977A&A....61...99B,1994A&A...284..331O}. 
Their flux densities and polarization properties are summarized in Table \ref{primecals}.

The individual flux density errors are composed of the statistical errors from the data 
reduction process (including the errors from the Gaussian fit, the weighted average over 
the sub-scans, and the gain and time-dependent corrections) and a contribution from the 
residual scatter seen in the primary and secondary calibrator measurements, which 
characterizes uncorrected residual effects. The total error in the relative flux density
measurements is typically $\Delta S/<S>\,\leq\,0.5-1$\,\%. At $\lambda$\,9\,mm, the increasing 
influence of the atmosphere (and larger pointing and focus errors) limit our calibration 
accuracy and the measurement error is increased by a factor of up to 2--3.

\subsection{Effelsberg: Linear Polarization}
The polarization analysis at 4.85 and 10.45\,GHz invokes correction of the measured 
Stokes Q/U channels for instrumental polarization, unequal receiver gains, cross-talk and 
depolarization. Since the measurements are done in the horizontal coordinate system (Alt-AZ), the 
parallactic angle rotation has to be taken into account. For this purpose we follow the polarization 
analysis using the matrix-method of \citet{1985A&A...142..181T} 
\citep[see also][for details]{1989A&A...226L...1Q,2003A&A...401..161K}. 
We obtained Stokes Q- and U-amplitudes by fitting Gaussian profiles to the pre-averaged 
sub-scans of each scan for the two orthogonal and 90 degree phase shifted receiver channels. 
The `true' flux density vector $S_{true}$ with the 3 components I, Q and U, which describe the source intrinsic 
polarization, is constructed using a 3$\times$3-M\"uller-matrix {\bf M}, 
\begin{eqnarray}
 S_{obs}={\bf M}\cdot S_{true}={\bf T}\cdot{\bf P}\cdot S_{true}\,\,.
\label{chi}
\end{eqnarray}
This allows the parameterization of the parallactic angle rotation via the time dependent 
matrix {\bf P} and the removal of the afore-mentioned systematic effects via inversion of the 
instrumental matrix {\bf T} (here we assume that Stokes V is negligible small; this reduces 
the matrix dimension to $3 \times 3$).

We used the polarization calibrator 3C286 and the highly polarized secondary calibrator 0836+710 
($p \simeq 6$\,\% to determine the elements of the instrumental matrix {\bf T} via a least-square 
fit procedure to $S_{obs}$, and by this the instrumental effects. In order to improve the accuracy 
of the determination of the instrumental polarization (of the order $<$\,0.5\,\%), the (non-variable) 
unpolarized secondary calibrator 0951+699 was also used. This procedure corrects for all (obvious) 
systematic effects and yields a relative measurement accuracy for the polarized flux density of 
$\frac{\Delta P}{P}\sim$\,5\% and $\Delta\chi=2-3^{\circ}$ for the polarization angle.   
\begin{table}
\begin{center}
\caption{Total and polarized flux densities and polarization
  angles for the calibrator sources at each observing frequency. 
  The quasar 0836+714 varies on longer time scales, but on IDV
  timescales is suitable as a secondary calibrator.}
\begin{tabular}{c|cccc}
\hline
\hline
Source   & $\lambda$ & $S_{I}$ & $S_{P}$ & $\chi$    \\
         & [mm]      & [Jy]    & [Jy]    & [$^\circ$]\\
\hline
3C286    & 60        & 7.49   &  0.82 & 33.05 \\      
         & 28        & 4.45   &  0.52 & 33.05 \\
         &  9        & 1.84   &    -- & --\\
3C295    & 60        & 6.53   &  0.01 & 71.54  \\
         & 28        & 2.64   &  0.04 & --23.61 \\
         &  9        & 0.55   &    -- & -- \\
3C48     & 60        & 5.50   & 0.23  & 106.77 \\
         & 28        & 2.59   & 0.15  & 116.36 \\
         &  9        & 0.80   &  --   & -- \\
NGC7027  & 60        & 5.47   & 0.01  & 55.61 \\  
         & 28        & 5.94   & 0.01  & 81.57 \\
         &  9        & 4.57   &   --  & -- \\ \hline
0836+710 & 60        & 2.54   & 0.16  & 108.13 \\
         & 28        & 2.12   & 0.09  & 103.37 \\
         &  9        & 1.82   &    -- &  -- \\
\hline
\hline
\end{tabular}
\label{primecals}
\end{center}
\end{table}

\subsection{Observations at higher frequencies}

At the IRAM 30\,m telescope, we followed a similar observing strategy as in Effelsberg.
In combination with the good telescope and receiver performance, a un-precedent precision of 
the mm-flux density light-curve was achieved, yielding an rms-accuracy of $\sim$\,1.2\,\% at 
86\,GHz \citep[see][for details]{2006A&A...456..117A}. At 229\,GHz the calibration accuracy was 
reduced due to the loss of 1 linear feed in the 1.3\,mm receiver (mal-function), which reduced 
the measurement accuracy to $\sim 16$\,\% rms at 229 GHz. This prevents a detailed study of short-term 
variability at this frequency, but still allows the measurement of the overall flux density trend.
%
%
   \begin{figure}
   \centering
   \vspace{0.5cm}
   \includegraphics[width=8.5cm]{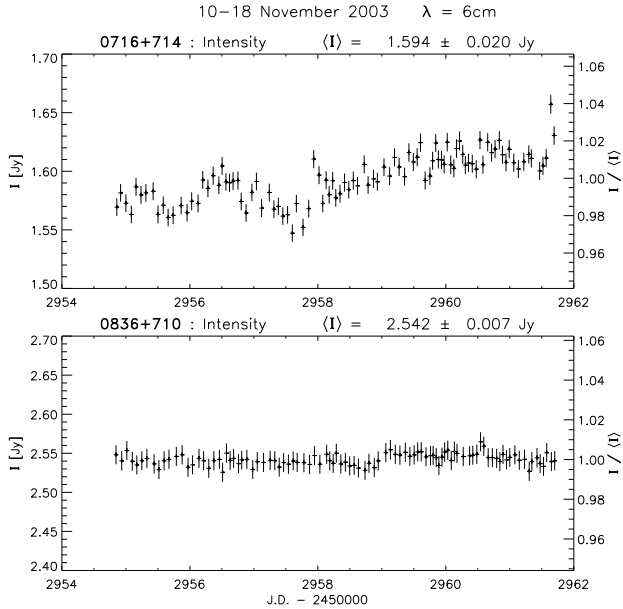}
      \caption{Flux density variability of 0716+714 (top panel) at 60\,mm
        wavelength. For comparison, the secondary calibrator 0836+710, which shows 
	residual peak-to-peak variations of $\sim$\,1\,\% ($m_{0}=0.3$\,\%), is 
	displayed (bottom panel).}
        \label{Fig_LC_6cm}
   \end{figure}
%
   \begin{figure}
   \centering
   \vspace{0.5cm}
   \includegraphics[width=8.5cm]{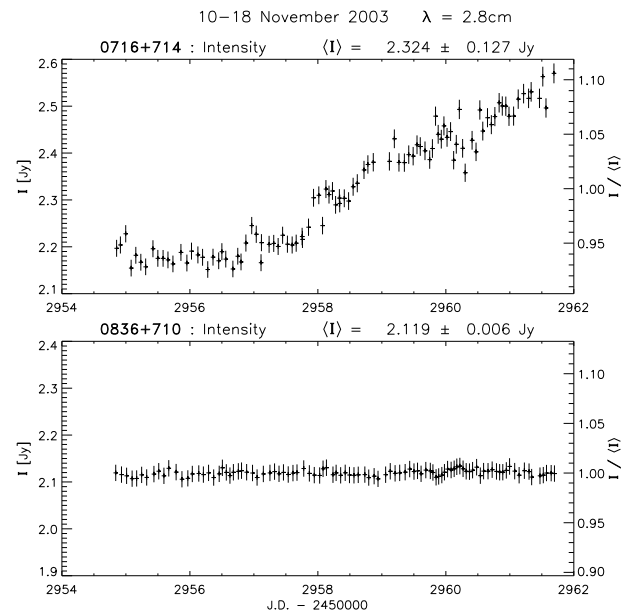}
	\caption{Flux density variability of 0716+714 (top panel) at 28\,mm wavelength.
	The secondary calibrator 0836+710 shows residual peak-to-peak variations of
	$\sim$\,1\,\% ($m_{0}=0.6$\,\%).}
         \label{Fig_LC_2.8cm}
   \end{figure}

The dense in time sampled monitoring at Effelsberg and Pico Veleta was complemented by
measurements with some smaller telescopes: Kitt Peak (3\,mm), SMT/HHT (0.87\,mm) and JCMT (0.85 \& 0.45\,mm).
The smaller telescope sizes and the correspondingly reduced sensitivity and the poorer time sampling
limit their data quality. For this reason, these data are mainly used to extend the spectral coverage
towards the sub-mm bands and to investigate the basic trend of the variability (see Sect. 3 and 4). A 
short summary of the observations with these telescopes is given below.

The measurements at Kitt Peak were performed using a SIS receiver operating 
at 90\,GHz, while the 345/666 GHz observations at the JCMT and the 345\,GHz observations 
at the SMT/HHT were performed using bolometer arrays. At the SMT/HHT the bolometer
consists of 19 individual continuum receivers arranged in concentric hexagons, allowing 
the subtraction of atmospheric effects using the off-horns \citep[][]{2002AIPC..616..262K}. 
At the JCMT, the continuum array receiver SCUBA was used, which has two hexagonal arrays of 
bolometric detectors with 37 and 91 horns, respectively \citep[][]{1999MNRAS.303..659H}.

At these three telescopes, the flux density of 0716+714 was measured using the on-source-off-source 
observing technique with the number of on/off subscans matched to the source brightness at the given 
frequency. At Kitt Peak and at the SMT/HHT the observing strategy was similar to the one used at 
Effelsberg and Pico Veleta (see Section \ref{obs} and \ref{eff_red}), consisting of frequent 
observations of primary and secondary calibrators (ultra-compact HII regions, planetary nebulae, 
planets, some AGN). Sky dips were done to allow for atmospheric opacity correction and the planets 
were used for the absolute flux density calibration (counts-to-Jansky conversion).
In all cases, the final measurement uncertainties are relatively high, up to 15--20\,\%, preventing 
us from studying the short-term variability pattern of 0716+714 at the highest frequencies. We 
therefore decided to improve the signal-to-noise ratio by averaging the individual measurements in 
24\,hr bins. This enables the characterization of the main flux density evolution at sub-mm wavelengths 
over the full observing period with a typical measurement accuracy of 5--10\,\%.    

%
   \begin{figure}
   \centering
   \vspace{0.5cm}
   \includegraphics[width=8.5cm]{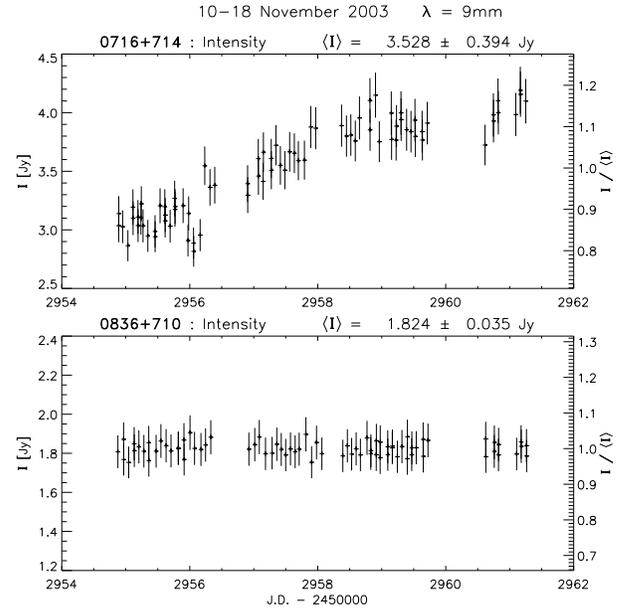}
      \caption{Flux density variability of 0716+714 (top panel) at 9\,mm wavelength.
              The secondary calibrator 0836+710 shows residual peak-to-peak variations of
	$\sim$\,8\,\% ($m_{0}=2.2$\,\%).}	
      \label{Fig_LC_9mm}
   \end{figure}
%
   \begin{figure}
   \centering
   \vspace{0.5cm}
   \includegraphics[width=8.5cm]{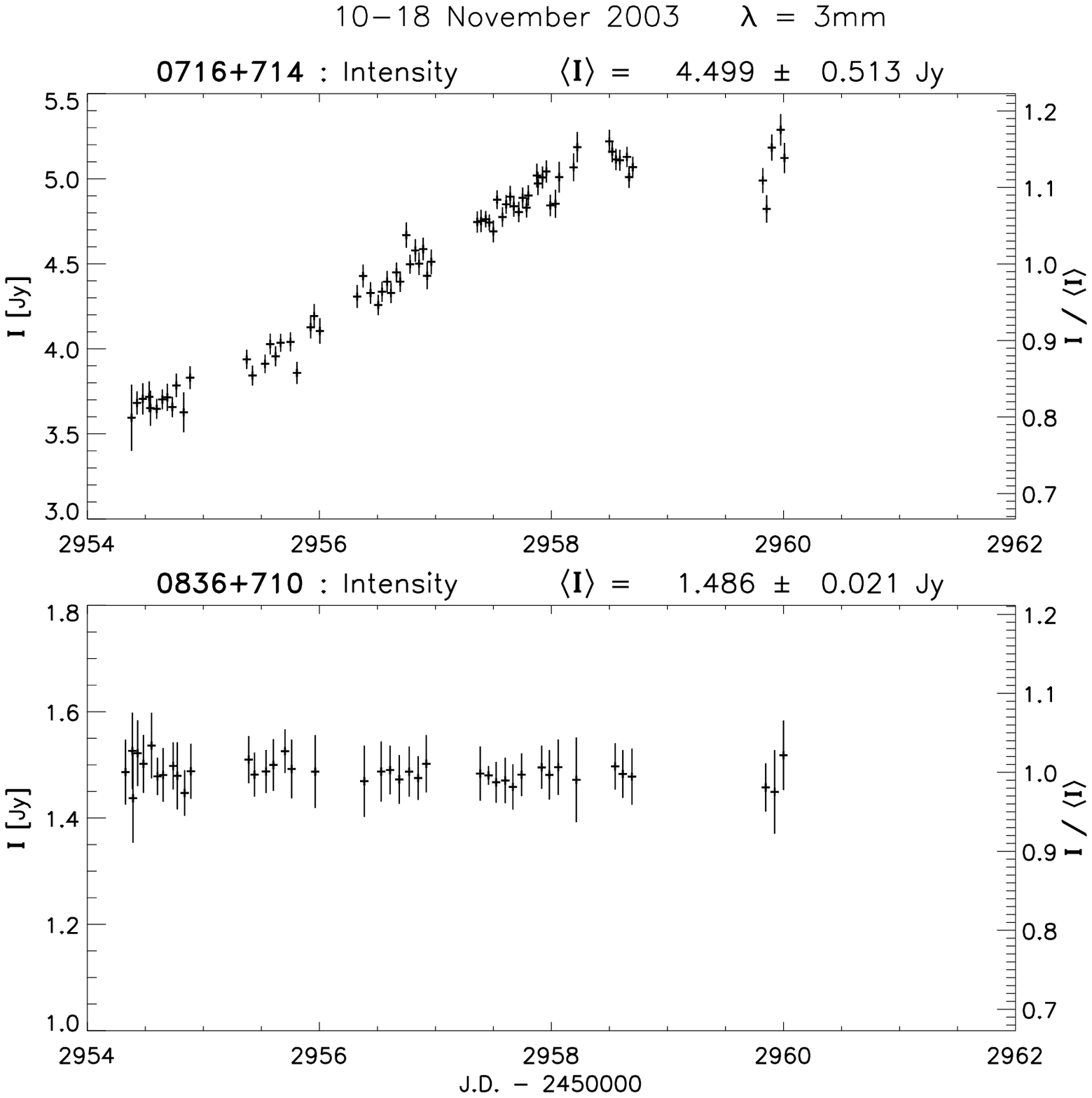}
	\caption{Flux density variability of 0716+714 (top panel) at 3.5\,mm wavelength.
	The secondary calibrator 0836+710 shows residual peak-to-peak variations of
	$\sim$\,6\,\% ($m_{0}=1.2$\,\%).}
         \label{Fig_LC_3mm}
   \end{figure}
%
   \begin{figure}
   \centering
   \vspace{0.5cm}
   \includegraphics[width=8.5cm]{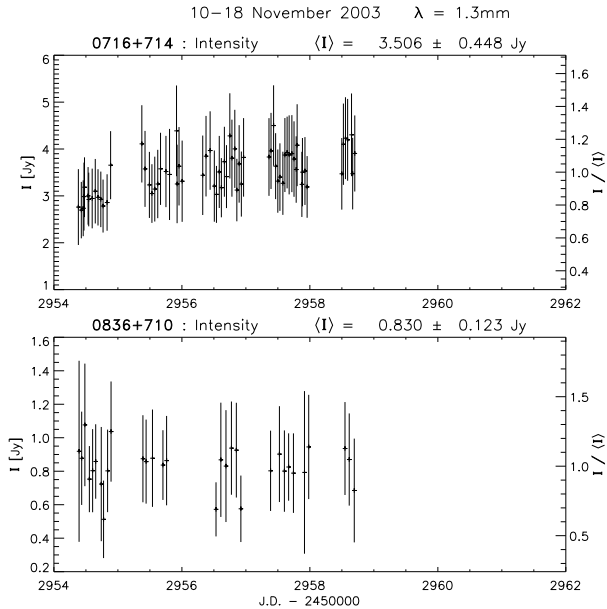}
	\caption{Flux density variability of 0716+714 (top panel) at 1.3\,mm wavelength. 
	The secondary calibrator 0836+710 shows residual peak-to-peak variations of
	$\sim$\,65\,\% ($m_{0}=16.0$\,\%) due to a receiver mal-function (see text).}	
         \label{Fig_LC_1mm}
   \end{figure}

\section{Data Analysis and Results}\label{analysis}

\subsection{Total intensity}\label{total_int}
In Fig. \ref{Fig_LC_6cm}, \ref{Fig_LC_2.8cm} and \ref{Fig_LC_9mm} we plot the
flux density measurements of 0716+714 versus time at 4.85, 10.45 and 32\,GHz. 
In Fig. \ref{Fig_LC_3mm} and \ref{Fig_LC_1mm} we report the measurements 
obtained with the IRAM 30\,m-telescope at 86 and 229\,GHz already published 
by \citet{2006A&A...456..117A}. For a direct comparison, the residual 
variability of the secondary calibrator 0836+714 is shown at each band.  

The light curves given in Fig. \ref{Fig_LC_6cm} to \ref{Fig_LC_3mm} show 0716+714 to be 
strongly variable, when compared to the stationary secondary calibrator. At $\lambda$\,60\,mm 
(Fig. \ref{Fig_LC_6cm}), 0716+714 exhibits low-amplitude variations with 
a slow ($\sim$\,2\,day) total flux density increase after J.D.\,$=$\,2452957.6 
(peak-to-peak variability amplitude $\geq 4$\,\%). 

The light curves obtained at 28, 9 and 3.5\,mm wavelengths (Fig. \ref{Fig_LC_2.8cm} to 
\ref{Fig_LC_3mm}) are dominated by a much stronger monotonic increase over a time range 
of several days between November 10 -- 18. In particular, we note an 
increasing amplitude of the variability towards higher frequencies ($\lambda$\,28\,mm: 
$\sim$\,16\,\% between J.D.\,$=$\,2452956.7 and J.D.\,$=$\,2452961.7; $\lambda$\,9\,mm: 
$\sim$\,25\,\% between J.D.\,$=$\,2452955.5 and J.D.\,$=$\,2452958.8). The most 
dramatic increase is seen at $\lambda$\,3\,mm (Fig. \ref{Fig_LC_3mm}). Here, 
the total flux density shows a linear increase of more than 35\,\% between 
J.D.\,$=$\,2452954.4 and J.D.\,$=$\,2452958.7. At $\lambda$\,1.3\,mm (Fig. \ref{Fig_LC_1mm}), 
a precise characterization of the temporal behavior is not possible due to the larger 
measurement errors. However, a linear fit to this data set indicates an increase
with a rate of change of 6\,\% per day \citep[see][for details]{2006A&A...456..117A}. 
     
In order to characterize the multi-frequency variability properties more quantitatively,
each data set was investigated by means of a statistical variability analysis based on 
the following steps: 
(i) a $\chi^{2}$-test for the presence of variability,
(ii) the measurement of the variability strength (modulation index $m$ and
noise-bias corrected variability amplitude $Y$), and (iii) the determination of
the characteristic variability time scales. These methods are described in more 
detail by \citet{1987AJ.....94.1493H}, \citet{1992A&A...258..279Q} and 
\citet{2003A&A...401..161K}. In the following we will use this formalism, and
refer to Appendix \ref{Ap1} for more details on the definitions of $\chi^{2}$, $m$, 
and $Y$. The determination of the variability time scales will be presented in Sect. 
\ref{var-analysis} and Appendix \ref{Ap2}.   
\begin{table*}
\begin{center}
\caption{Results of the variability analysis for total intensity. N denotes the
  total number of measurements, $<S_{I}>$ the mean flux density with the rms-value $\sigma_S$,
  $m$ the modulation index, $Y$ the variability amplitude, $\chi^{2}_{r}$ the reduced $\chi^{2}$ 
  and $\chi^{2}_{99.9\%}$ the corresponding value for a 99.9\,\% significance level of variability 
  (see Appendix \ref{Ap1} for definitions).} 
\begin{tabular}{r|rrrrrrr}
\hline
\hline
\multicolumn{8}{c}{S5\,0716+714}\\
\hline
\hline
$\lambda$ & N   & $<S_{I}>$ & $\sigma_S$ & $m$   & $Y$   & $\chi^{2}_{r}$ & $\chi^{2}_{99.9\%}$ \\
\,[mm]    &     & [Jy]      & [Jy]       & [\%]  & [\%]  &                &                    \\
\hline
60        & 100 & 1.594     & 0.020      & 1.28  & 3.72  & 7.12           & 1.50\\
28        & 98  & 2.324     & 0.127      & 5.45  & 16.26 & 43.87          & 1.50\\
9         & 79  & 3.528     & 0.391      & 11.09 & 32.73 & 6.38           & 1.57\\
3.5       & 74  & 4.502     & 0.516      & 11.46 & 34.19 & 55.30          & 1.59\\
1.3       & 68  & 3.506     & 0.448      & 12.77 & $-$   & 0.50           & 1.62\\
\hline
\hline
\end{tabular}
\end{center}
\begin{center}
\begin{tabular}{l|rrrrrrrr}
\hline
\hline
\multicolumn{9}{c}{Secondary calibrators}\\
\hline
\hline
\multicolumn{9}{l}{$\lambda$\,60mm~~~$m_{I,0}=0.30$\,\%,~~~~$\lambda$\,28mm~~~$m_{I,0}=0.55$\,\%,~~~~$\lambda$\,9mm~~~$m_{I,0}=2.18$\,\%}\\
\multicolumn{9}{l}{$\lambda$\,3.5mm~~$m_{I,0}=1.20$\,\%,~~~~$\lambda$\,1.3mm~~$m_{I,0}=16.0$\,\%}\\
\hline
Source   & $\lambda$ & N & $<S_{I}>$ & $\sigma_S$ & $m$ & $Y$  & $\chi^{2}_{r}$ & $\chi^{2}_{99.9\%}$ \\
         & [mm]      &   & [Jy]  & [Jy]     & [\%]& [\%] &                &                    \\
\hline
0212+735 & 3.5       & 33  & 1.101 & 0.016 & 1.45  & $-$   & 0.31  & 1.95\\    
         & 1.3       & 15  & 0.712 & 0.094 & 13.14 & $-$   & 0.12  & 2.58\\
\hline
0633+734 & 3.5       & 25  & 0.942 & 0.018 & 1.90  & $-$   & 0.40  & 2.13\\
         & 1.3       & 15  & 0.621 & 0.085 & 13.72 & $-$   & 0.11  & 2.58\\
\hline
0633+599 & 28        & 56  & 0.677 & 0.004 & 0.56  & $-$   & 0.46  & 1.69\\
         & 9         & 51  & 0.639 & 0.015 & 2.42  & $-$   & 0.28  & 1.73\\
\hline
0800+618 & 60        & 94  & 1.385 & 0.004 & 0.31  & $-$   & 0.42  & 1.52\\
         & 28        & 96  & 1.130 & 0.007 & 0.61  & $-$   & 0.54  & 1.51\\
\hline
0835+583 & 28        & 57  & 0.239 & 0.002 & 0.71  & $-$   & 0.64  & 1.69\\
\hline
0836+710 & 60        & 98  & 2.542 & 0.007 & 0.28  & $-$   & 0.34  & 1.50\\
         & 28        & 99  & 2.119 & 0.006 & 0.30  & $-$   & 0.13  & 1.50\\
         & 9         & 75  & 1.824 & 0.035 & 1.93  & $-$   & 0.18  & 1.59\\
         & 3.5       & 42  & 1.486 & 0.020 & 1.36  & $-$   & 0.14  & 1.82\\
         & 1.3       & 31  & 0.830 & 0.123 & 14.77 & $-$   & 0.28  & 1.99\\
\hline
0951+699 & 60        & 48  & 3.298 & 0.008 & 0.25  & $-$   & 0.27  & 1.75\\
         & 28        & 46  & 1.812 & 0.010 & 0.58  & $-$   & 0.51  & 1.78\\
\hline
1803+784 & 3.5       & 32  & 1.334 & 0.016 & 1.19  & $-$   & 0.09  & 1.97\\
         & 1.3       & 20  & 0.940 & 0.235 & 25.04 & $-$   & 0.78  & 2.31\\
\hline
1928+738 & 3.5       & 32  & 1.866 & 0.024 & 1.31  & $-$   & 0.18  & 1.97\\
         & 1.3       & 20  & 0.904 & 0.177 & 19.54 & $-$   & 0.44  & 2.31\\
\hline
\hline
\end{tabular}
\label{results_I_table}
\end{center}
\end{table*}

In Table 4 we summarize the results of the variability analysis of the total flux 
density measurements for 0716+714 (and the observed secondary calibrators) at the 
five observing wavelengths. Here, for each source and each wavelength, the number 
of data points N, the mean flux density $<S>$ and the modulation index $m$ is given. 
The (weighted) mean modulation index $m_{0}$ of the secondary calibrators is shown 
in the header of Table 4 for each frequency. The corresponding values of $m_0=$ 0.3, 0.6, 2.2 
and 1.2\,\% obtained at $\lambda= 60, 28, 9, and 3$\,mm demonstrate the small residual scatter in 
the secondary calibrator data and thus the good overall calibration accuracy.
%
   \begin{figure*}
   \centering
   \includegraphics[width=16.5cm,angle=0]{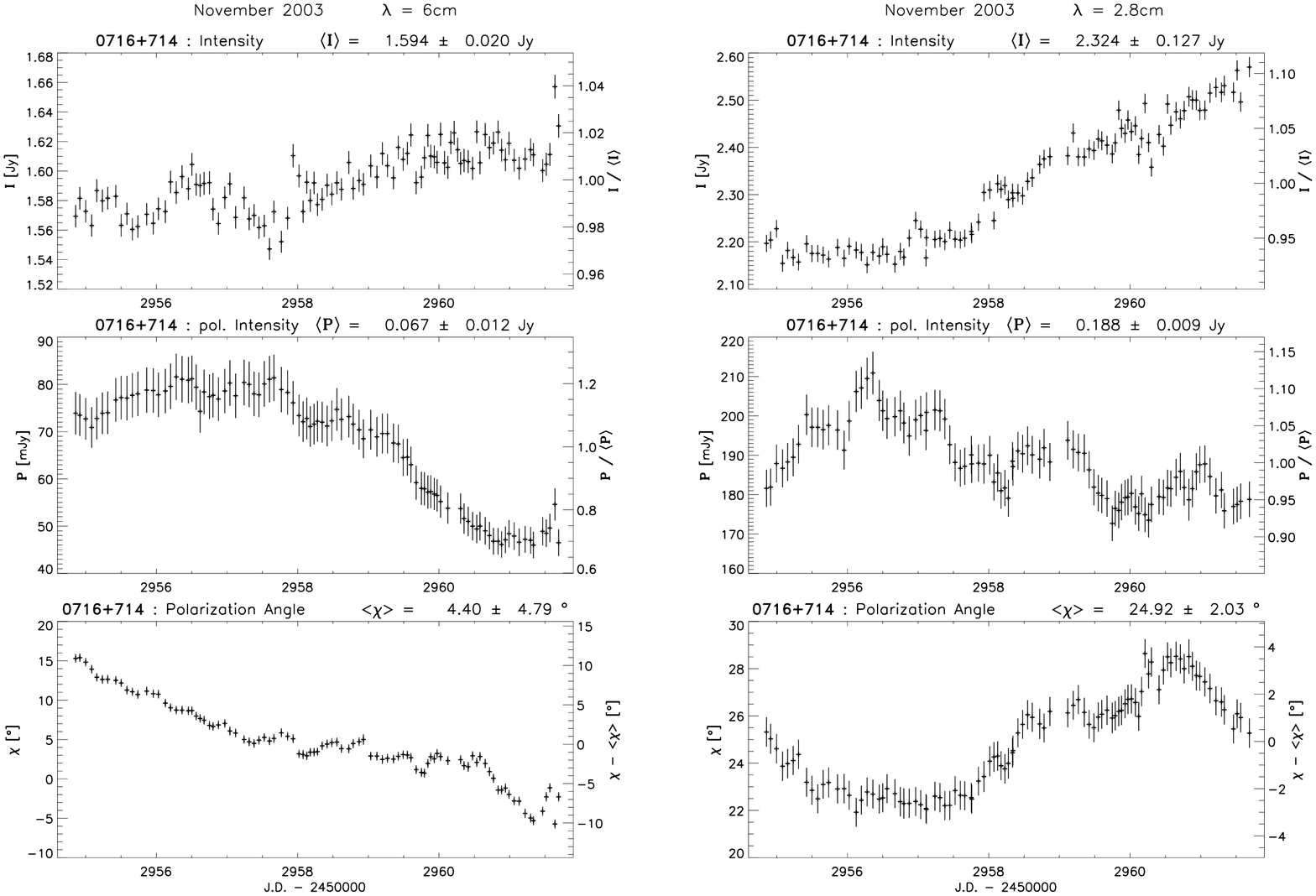}
      \caption{Flux density variability of 0716+714 obtained at 60\,mm (left panel) and 28\,mm 
	wavelength (right panel) in total intensity (top), polarized intensity (middle), and 
	polarization angle (bottom), respectively.}
         \label{Fig_LC_pol_6+2.8cm}
   \end{figure*}
%

The results of the $\chi^{2}$-test for each data set are given in terms of the reduced 
$\chi_{r}^{2}$ and the corresponding value at which a 99.9\,\% significance level for 
variability is reached ($\chi^{2}_{99.9\%}$). The variability amplitude Y is only 
calculated for those sources which according to the $\chi^{2}$-test, showed significant 
variability. It is obvious from Table 4, that only 0716+714 shows significant variability at
$\lambda$\,60, $\lambda$\,28, $\lambda$\,9 and $\lambda$\,3\,mm.

\subsection{Polarization}

In Fig. \ref{Fig_LC_pol_6+2.8cm} we show the
polarization variability of 0716+714 measured with the 100\,m telescope at 
$\lambda$\,60\,mm and $\lambda$\,28\,mm. Here, the time evolution of the total 
flux density $I$ is displayed together with the polarized intensity $S_{P}$ and the
polarization angle $\chi$. For a detailed discussion of the polarization data 
obtained at Pico Veleta we refer to \citet{2006A&A...456..117A}. Due to the marginal 
($2\sigma$) detection of polarization variations at 3\,mm wavelength, we do 
not consider this in the further analysis.   

A first inspection of the linear polarization data shown in
Fig. \ref{Fig_LC_pol_6+2.8cm} reveals for 0716+714 strong and rapid 
variability also in the polarized intensity and the polarization 
angle. In order to characterize the variability behavior in polarization, a
variability analysis similar to that described for the total intensity in
Sect. \ref{total_int} was performed (see Appendix \ref{Ap1}). The results
of the corresponding quantities ($m_{P}$, $Y_{P}$) for the polarization
$S_{P}$ are summarized in Table 5. A $\chi^{2}$-test was performed to 
determine the significance of the polarization variability. The corresponding 
values of $\chi^{2}_{r}$ and $\chi^{2}_{99.9\%}$ for $S_{P}$ and $\chi$ 
are also included in the Table 5. 

The statistical variability test shows significant and strong variability of 0716+714 
also in polarized flux and polarization angle. At $\lambda$\,60\,mm, the polarized 
intensity displays a $10-15$ times higher variability amplitude than the total 
intensity variations. The pronounced decrease of the polarized flux occurring 
after J.D.\,$=$\,2452956.5 (November 13), seems to anti-correlate with the 
increase of the total flux density observed during the same time interval. 
The polarization angle, however, changed nearly monotonically by about 
20$^\circ$ over the whole observing period. The linear polarization data obtained 
at $\lambda$\,28\,mm exhibit an overall variability amplitude lower than that 
of the $\lambda$\,60mm data. At $\lambda$\,28\,mm, the relative strength of 
the variations seen in the polarized and total flux are comparable, but the 
polarization variations appear faster. We further note that the amplitude of 
the polarization angle variations at $\lambda$\,28\,mm are about a factor of 4 
lower than at $\lambda$\,60\,mm. Similar as at $\lambda$\,60\,mm, the polarized flux 
density variations appear to be anti-correlated with the total intensity also at
$\lambda$\,28\,mm. This will be further investigated in Sect. \ref{dcfs}.

\begin{table*}
\begin{center}
\caption{Results of the variability analysis for polarization. N denotes the total number of 
measurements, $<S_{P}>$ the mean polarized intensity, $\sigma_{S_P}$ its rms-value, $m_{P}$ the modulation 
index, $Y_{P}$ the variability amplitude, $<\chi>$ the mean polarization angle, $\sigma_{\chi}$ its rms-value. 
The reduced $\chi^{2}_{r}$ is given for polarized intensity and polarization angle. $\chi^{2}_{99.9\%}$ denotes 
the corresponding 99.9\,\% significance level for variability. See Appendix \ref{Ap1} for definitions.}
\begin{tabular}{r|rrrrrrrrrr}
\hline
\hline
\multicolumn{11}{c}{S5\,0716+714}\\
\hline
\hline
$\lambda$ & N & $<S_{P}>$ & $\sigma_{S_P}$ & $m_{P}$ & $Y_{P}$ & $\chi^{2}_{r}$ & $<\chi>$   & $\sigma_{\chi}$ & $\chi^{2}_{r}$ & $\chi^{2}_{99.9\%}$\\
\,[mm]    &   & [Jy]      & [Jy]           & [\%]    &   [\%]  &                & [$^\circ$] &  [$^\circ$]     &       
&  \\
\hline
60        & 98 & 0.067   & 0.012 & 18.14 &  54.42 & 12.86 &   4.40 &  4.79 &  77.25 & 1.50 \\
28        & 98 & 0.188   & 0.009 &  4.78 &  14.33 &  0.35 &  24.92 &  2.03 &  10.85 & 1.50 \\
\hline
\hline
\end{tabular}
\end{center}
\begin{center}
\begin{tabular}{l|rrrrrrrrrrr}
\hline
\hline
\multicolumn{12}{c}{Polarization calibrators}\\
\hline
\hline
\multicolumn{5}{l}{$\lambda$\,60mm~~~~~~~~~~~$m_{P,0}=1.8$\,\%~~$\sigma_{\chi,0}=0.6^\circ$} \\
\multicolumn{5}{l}{$\lambda$\,28mm~~~~~~~~~~~$m_{P,0}=2.0$\,\%~~$\sigma_{\chi,0}=0.5^\circ$} \\
\hline
\hline
Source&$\lambda$&N&$<S_{P}>$&$\sigma_{S_P}$& $m_{P}$&$Y_{P}$&$\chi^{2}_{r}$&$<\chi>$&$\sigma_{\chi}$&$\chi^{2}_{r}$
&$\chi^{2}_{99.9\%}$\\
         & [mm]  &   & [Jy] & [Jy] & [\%] & [\%]  &   & [$^\circ$] &  [$^\circ$]     &                &  \\
\hline
3C286    & 60        & 16 & 0.826   & 0.009 &  1.08 & $-$ & 0.28 & 33.02 & 0.19 & 0.31 & 2.51 \\
         & 28        & 15 & 0.521   & 0.008 &  1.45 & $-$ & 0.51 & 32.95 & 0.16 & 0.09 & 2.58 \\
\hline
0836+710 & 60        & 95 & 0.156   & 0.002 &  1.79 &   $-$  &  0.51 & 108.17 &  0.29 &   0.76 & 1.51 \\
         & 28        & 97 & 0.085   & 0.002 &  1.97 &   $-$  &  0.84 & 103.56 &  0.57 &   1.08 & 1.51 \\
\hline
0951+699 & 60        & 48 & 0.003   & 0.002 & $-$ & $-$ &  $-$ & $-$ & $-$ & $-$ & $-$ \\
         & 28        & 42 & 0.002   & 0.001 & $-$ & $-$ &  $-$ & $-$ & $-$ & $-$ & $-$ \\
\hline
\hline
\end{tabular}
\label{results_pol}
\end{center}
\end{table*}

\subsection{Variability characteristics}\label{var-analysis}

\subsubsection{Variability time scales for total intensity and polarization}
To further quantify the variability behavior of 0716+714, characteristic variability 
time scales have been extracted from the light curves shown in Fig. 
\ref{Fig_LC_6cm}--\ref{Fig_LC_pol_6+2.8cm}. The light curves are dominated by a 
quasi-monotonic flux density increase over several days (and a decrease in the polarization), 
with occasional super-imposed less pronounced but more rapid variations. The pronounced, 
quasi-periodic rapid ($< 1$\,day time scale) variability, which is usually seen in this 
and other IDV sources \citep[e.g.][]{1995ARA&A..33..163W,2003A&A...401..161K}, is 
not really obvious in this campaign. Therefore, a precise determination of a 
characteristic variability time scale (in a statistical sense) is more difficult. 
We will therefore use a number of different methods to determine the variability time 
scales: structure function (SF), auto-correlation function (ACF) and a minimum-maximum 
method. These methods and the corresponding error calculations are described in more detail 
in Appendix \ref{Ap2} and the results will be presented in the following. We confine 
our variability analysis for 0716+714 to the most accurately measured data sets and 
thus exclude the light curve obtained at 1.3\,mm wavelength. In Table \ref{timescales} 
we summarize the derived variability time scales and their measurement errors for each 
of the different data sets.  
%
   \begin{figure*}
   \centering
   \includegraphics[width=18cm,angle=0]{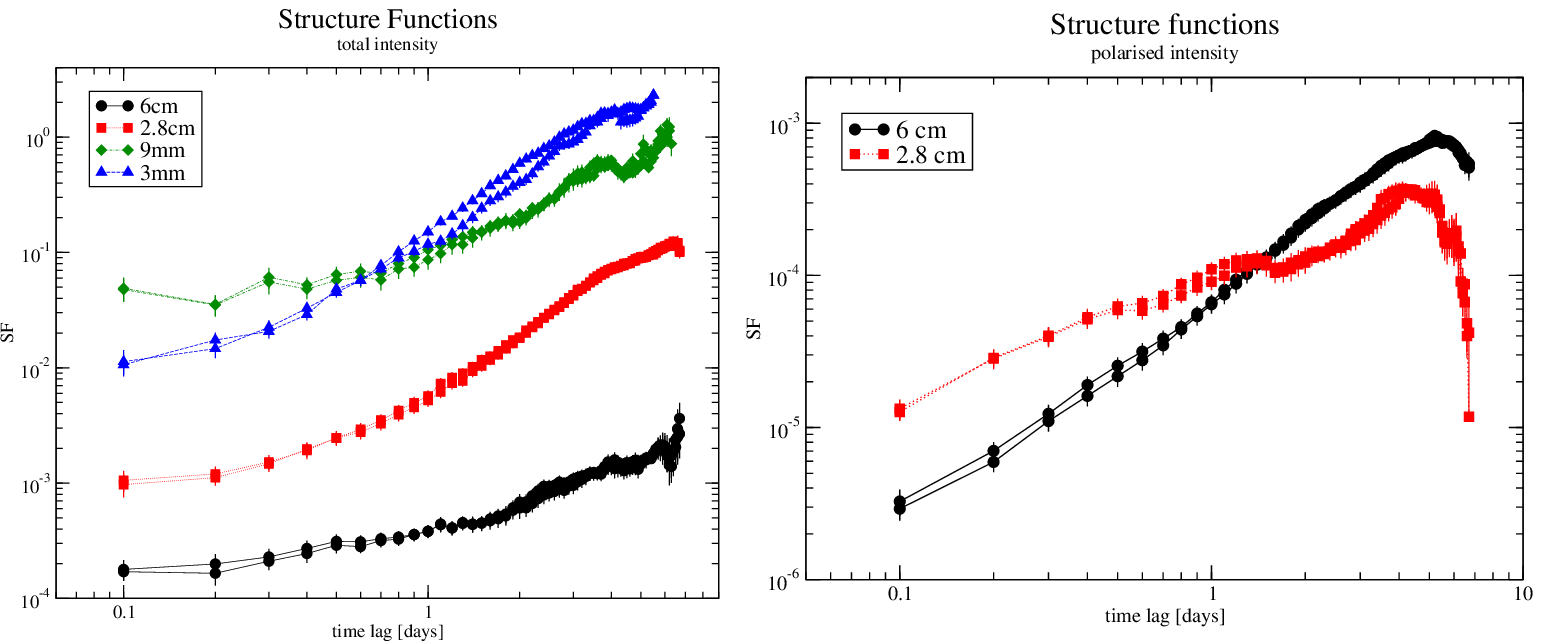}
      \caption{Structure functions of 0716+714 plotted versus time-lag. 
        Different symbols are used for different wavelengths. The panel
        on the left shows the SF for total intensity, on the right the SF for polarized intensity
        is shown.}
        \label{SFs1}
   \end{figure*}
%

In Fig. \ref{SFs1} we show examples of calculated structure functions for the total 
and polarized flux density. We note that the error of the derived time scales from the SF and 
ACF analysis shown in Table \ref{timescales} is usually high - values of up 
to 30\,\% were obtained. The uncertainty of the statistical parameters which quantify
the variations in a time series critically depends on the duration of the observation, 
the sampling interval and the number of significant flux density changes 
(variability cycles) observed. The light curves of 0716+714 
(Fig. \ref{Fig_LC_6cm}--\ref{Fig_LC_pol_6+2.8cm}) often show only a monotonic increase 
and a well defined variability time scale is not observed. This allows only to derive 
lower limits to the variability time scales (see Appendix \ref{Ap2}). In the other cases 
of better defined variability time scales, the measured accuracy is a few 10\,\%. Here 
the values obtained by the three different analysis methods are in good agreement.

The prominent and monotonic flux density increase seen at 60, 28, 9 and 3\,mm wavelengths 
appears on similar time scales. The SF analysis yields lower limits of 3.7 to 4.2\,days 
(ACF: 3.0--3.7\,days). Due to the lack of pronounced saturation levels in the SFs
on time scales $\le$\,2\,days (see Fig. \ref{SFs1}), we identify the observed inter-day 
variability of 0716+714 in total intensity as IDV of type-I according to \citet{1987AJ.....94.1493H}. 
A significant component of faster variability - with a time scale of $<$\,1\,day - was 
found only at 60mm wavelength. The increasing amplitudes of the SFs shown in Fig. 
\ref{SFs1} (left), nicely demonstrate the increasing strength of the observed variations 
towards higher frequencies. The value of the SF at small time lags, $\tau\rightarrow 0$, 
characterizes the noise level present in the time series. We note that for small time 
lags ($\tau < 0.7$\,days), the value of the $\lambda$\,9\,mm SF in Fig. \ref{SFs1} (top) 
is larger than the $\lambda$\,3\,mm SF. This can be interpreted with a reduced short time 
stability of the 9\,mm signal, which is affected by the stronger atmospheric influence at 
the Effelsberg site compared to the high-altitude Pico Veleta site.
%
\begin{table*}
\begin{center}
\caption{Variability time scales and lower limits to the variability brightness temperatures 
for total intensity and polarization in 0716+714. The variability brightness temperatures were derived 
using Eq. (\ref{T_b}).}
\begin{tabular}{c|lcclccccc}
\hline
\hline 
\multicolumn{10}{c}{total intensity}\\
\hline
$\lambda$ & $t_{SF}$ & $\delta t_{SF}$ & $T_{B}$ & $t_{ACF}$ & $\delta t_{ACF}$ & $T_{B}$ & $\Delta t$ & $t_{var}$ & $T_{B}$\\
~[mm]     & [days]   & [days]    &  [K]                  & [days]    & [days] & [K]     & [days]  & [days]     & [K]\\
\hline
60        & 0.9      & +0.2/--0.2  & 3.7$\cdot10^{16}$ & 0.6      & +0.1/--0.1 & 8.1$\cdot10^{16}$ & --   & -- & -- \\
          & 4.2      & +0.4/--0.4  & 2.7$\cdot10^{15}$ & 3.1      & +0.4/--0.4 & 4.9$\cdot10^{15}$ & --   & -- & --\\
28        & $>$\,4.3 & +0.3/--0.3  & --                & $>$\,3.7 & +0.2/--0.3 & --                & 4.6 & 20.9 & 4.0$\cdot10^{14}$\\
 9        & $>$\,3.7 & +0.4/--0.5  & --                & $>$\,3.2 & +0.2/--0.3 & --                & 3.5 & 10.6 & 4.2$\cdot10^{14}$\\
 3.5      & $>$\,3.8 & +0.1/--0.2  & --                & $>$\,3.0 & +0.2/--0.2 & --                & 4.1 & 10.7 & 6.5$\cdot10^{13}$\\
\hline
\multicolumn{10}{c}{linear polarization}\\
\hline
60        & $>$\,4.5 & +0.2/--0.2 & --                & $>$\,3.2 & +0.2/--0.2 & --                & 4.6   & 8.16 & 5.6$\cdot10^{14}$ \\
28        & 1.3      & +0.2/--0.2 & 2.1$\cdot10^{15}$ & --       & --         & --                &-- &-- &-- \\ 
          & 3.8      & +0.3/--0.4 & 3.2$\cdot10^{14}$ & 2.8      & +0.5/--0.6 & 5.6$\cdot10^{14}$ &-- &-- &-- \\
\hline
\multicolumn{9}{c}{polarization angle}\\
\hline
60        & $>$\,6.8 & --              & -- & $>$\,6.8 & --           & -- &-- &-- &-- \\
28        & 3.8      & +0.1/--0.2    & -- & 3.1     & +0.2/--0.2 & -- &-- &-- & --\\
\hline
\hline
\end{tabular}
\label{timescales}
\end{center}
\end{table*}

In linear polarization, the source varies at $\lambda$\,60\,mm on time scales comparable
to those seen in total intensity, whereas at $\lambda$\,28\,mm, two variablity
time scales are seen: one comparable to the time scale of the
total intensity variations ($\sim 4$\,days) and a significantly faster
component, showing $t_{var}=1.3$\,days. The polarization angle changes at 
$\lambda$\,60\,mm display a continuous decrease over the seven days and our 
analysis yields a lower limit of $t_{var} \geq 6.8$\,days. In contrast, the 
variations of the polarization angle at $\lambda$\,28\,mm appear faster by a 
factor of about two.
%
   \begin{figure*}
   \centering
   \includegraphics[width=18cm,angle=0]{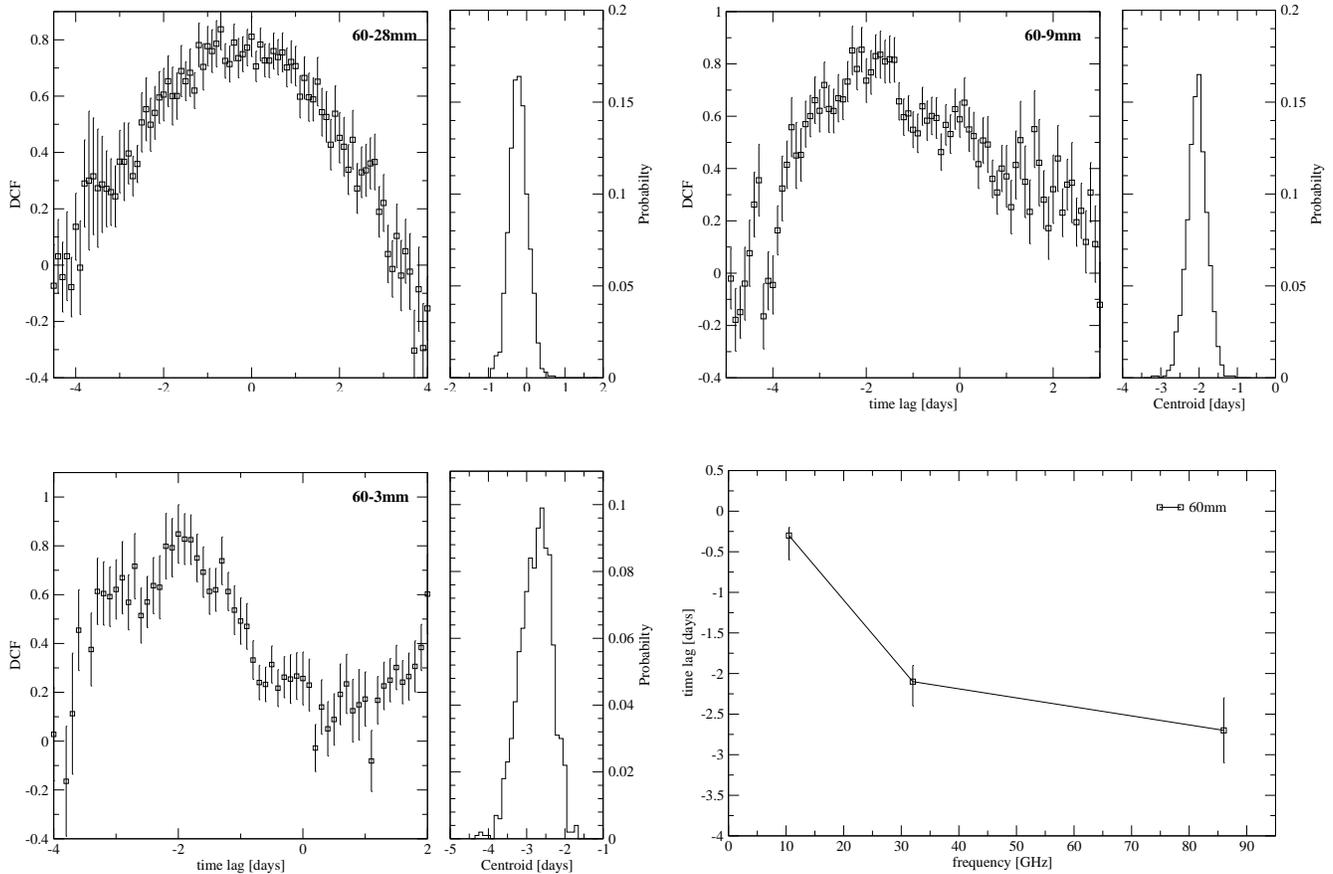}
      \caption{Examples of cross-correlation functions between total intensity light 
	curves obtained at the different frequencies. Here, we show combinations with 
        respect to the data at 60\,mm (see top right labels in each plot). The CCPD for the 
	peak of the DCF is shown in the right-hand panels (for details see text). In the bottom 
	right panel, we plot the time lags versus frequency, relative to the light curve at 60\,mm. 
        Note the trend of increasing (negative) time lag with increasing frequency.}
         \label{ccfs}
   \end{figure*}
  
\subsubsection{Variability time scales and brightness temperatures}\label{Tb-calc}
Assuming that the observed rapid variability would be source intrinsic (see Sect. 
\ref{ISS_mod}), the variability time scales summarized in Table \ref{timescales} 
imply very compact emission regions, and hence very high intrinsic brightness temperatures.
Following \citet{1979ApJ...233..498M}, and taking an isotropically
expanding light-sphere without preferred direction into account, the light
travel time argument implies a diameter $d\le2\cdot\,c\Delta\,t$ of the emitting region 
after the time interval $\Delta t$ [s] of expansion. We then obtain for the
diameter of the variable emission region $\theta$ [mas]:
\begin{eqnarray}
\theta=0.13\,\frac{t_{var}}{d_{L}}\,\delta(1+z)\,\,,
\label{dim}
\end{eqnarray}
where $z$ is the redshift of the source, $\delta$ the Doppler factor, $t_{var}$
the logarithmic variability time scale in years (see Appendix \ref{Ap2}, eq. B5) and 
$d_{L}$ the luminosity distance in Gpc. Consequently, we find for the brightness temperature 
$T_{B}$ [K] of a stationary component with a Gaussian brightness distribution 
\citep[e.g.][]{2005AJ....130.2473K} and flux density $S_{\lambda}$ [Jy]:
\begin{eqnarray}
T_{B}=8.47\times10^{4}\cdot S_{\lambda}\left(\frac{\lambda\,d_{L}}{t_{var,\lambda}\,(1+z)^{2}}\right)^{2}\,\,.
\label{T_b}
\end{eqnarray}
According to Eq. (\ref{T_b}) we calculated the brightness temperatures for the 
`most reliably determined' variability time scales derived in the previous section 
(see also Appendix \ref{Ap2}). Here and in the following we adopt a cosmological 
distance corresponding to $z=0.3$, which is a lower limit for the redshift of the source 
{\bf (see Sect. \ref{intro})}, and yields a luminosity distance of $d_{L}=1.51$\,Gpc assuming 
a flat universe with $H_{0}=72$\,km\,s$^{-1}$\,Mpc$^{-1}$, $\Omega_{M}=0.3$, $\Omega_{\Lambda}=0.7$ 
and $\Omega_{k}=0$ \citep[][]{2003ApJS..148..175S}. In Table \ref{timescales}, we summarize 
the calculated values of $T_{B}$ for each observing band and for total and polarized 
intensity, respectively. 

For the slow variability of type I seen in the cm-bands, the brightness temperatures
range between $4.0\!\cdot\!10^{14}$\,K and $4.9\!\cdot\!10^{15}$\,K, whereas in 
the mm-bands, we obtain lower values ranging between $6.5\,\cdot\,10^{13}$\,K
and $4.2\,\cdot\,10^{14}$\,K. The faster type II variability component seen 
at $\lambda\,60$\,mm, leads to a brightness temperature of up to $8.1\,\cdot\,10^{16}$\,K, 
which is at least one order of magnitude higher than the previous estimates. The 
variations observed in polarized intensity reveal brightness temperatures ranging 
between $3.2\,\cdot\,10^{14}$\,K and $2.1\,\cdot\,10^{15}$\,K, similar to those 
obtained for total intensity in the cm-band. We note that in our calculations 
we used conservative estimates of the variability time scales, which yield 
hard lower limits to $T_B$. Depending on the assumptions on the geometry and isotropy 
of the emitting region, the actual brightness temperatures may be higher by 
a factor of up to 6 (e.g. assuming for the emission region a uniform flat disk 
of size $d\simeq c\Delta t$ as in the case of a shock). We also note that a 
larger cosmological distance of 0716+714 would further increase $T_{B}$.  

\subsection{Cross correlations}\label{dcfs}
To further quantify the close similarities seen in the light curves of 0716+714 
across the observing bands, we computed the discrete cross-correlation
function (DCF) between different frequencies allowing to search for correlations and 
possible time lags. We followed the method described by \citet{1988ApJ...333..646E} 
and \citet{1992ApJ...386..473H} for unevenly sampled data, and calculated the DCF 
and the position of its maximum by using the centroid $\tau_{c}$ of the DCF, 
given by
\begin{eqnarray}
\tau_c=\frac{\sum_{i}{\tau_{i}DCF_{i}}}{\sum_{i}DCF_{i}}\,\,.
\label{centr}
\end{eqnarray}
In order to obtain statistically meaningful values for the cross-correlation
time lags and their related uncertainties, we performed Monte Carlo simulations, 
following the methods described in detail by \citet{1998PASP..110..660P} and 
\citet{2003A&A...402..151R}. Here, we take into account the influence of both uneven 
sampling and flux density errors. This was done using random subsets of the
two data sets. In addition, random Gaussian fluctuations constrained by the 
measurement errors were added to the flux densities. In each simulation we determined 
the centroid $\tau_{c}$ of the DCF peak. After running 1000 simulations, we obtained 
a cross-correlation peak distribution (CCPD), which is shown in the right panel of 
Fig. \ref{ccfs}. This technique yields a reliable measure of the uncertainties in 
the estimated time lags whereas the values, computed directly from the CCPDs, 
correspond to 1$\sigma$ errors \citep[see][]{1998PASP..110..660P}. 

This procedure was applied to each possible frequency combination of the total 
intensity data (60/28\,mm, 60/9\,mm, 60/3\,mm, 28/9\,mm, 28/3\,mm, 9/3\,mm). 
Fig. \ref{ccfs} shows some examples of the DCFs and CCPDs relative to the
$\lambda\,60$\,mm data. Our analysis confirms the existence of a significant 
correlation across all observed radio-bands. This enables the determination 
of time lags, which are $\tau_{60/28}\!=\!-0.3^{+0.1}_{-0.3}$\,days 
between the 60\,mm and the 28\,mm data, $\tau_{60/9}\!=\!-2.1^{+0.2}_{-0.3}$\,days 
between the 60\,mm and the 9\,mm data, and $\tau_{60/3}\!=\!-2.7^{+0.4}_{-0.4}$\,days 
between the 60\,mm and the 3\,mm data. This systematic trend is also seen in the time 
lags relative to the 28\,mm data, with $\tau_{28/9}\!=\!-1.9^{+0.2}_{-0.2}$\,days and 
$\tau_{28/3}\!=\!-2.4^{+0.2}_{-0.2}$\,days, respectively. In Fig. \ref{ccfs} (bottom right) 
we plot the $\tau_{60/i}$ time lags for the 3 observing bands ($i=28, 9, 3$\,mm) versus 
observing frequency. A systematic trend of increasing (negative) time lag towards higher
frequencies is evident. This shows that the observed variations (the observed flux 
density increase) occur first at higher frequencies and then propagate through the 
radio spectrum towards lower frequencies. The time delay between the two most 
separated bands ($\lambda\,3$\,mm and $\lambda\,60$\,mm) is about 2.5\,days.
This time-lag behavior together with the observed increasing variability amplitudes 
will be discussed in Sect. 4. 

\begin{figure}
   \centering
   \includegraphics[width=7.5cm,angle=0]{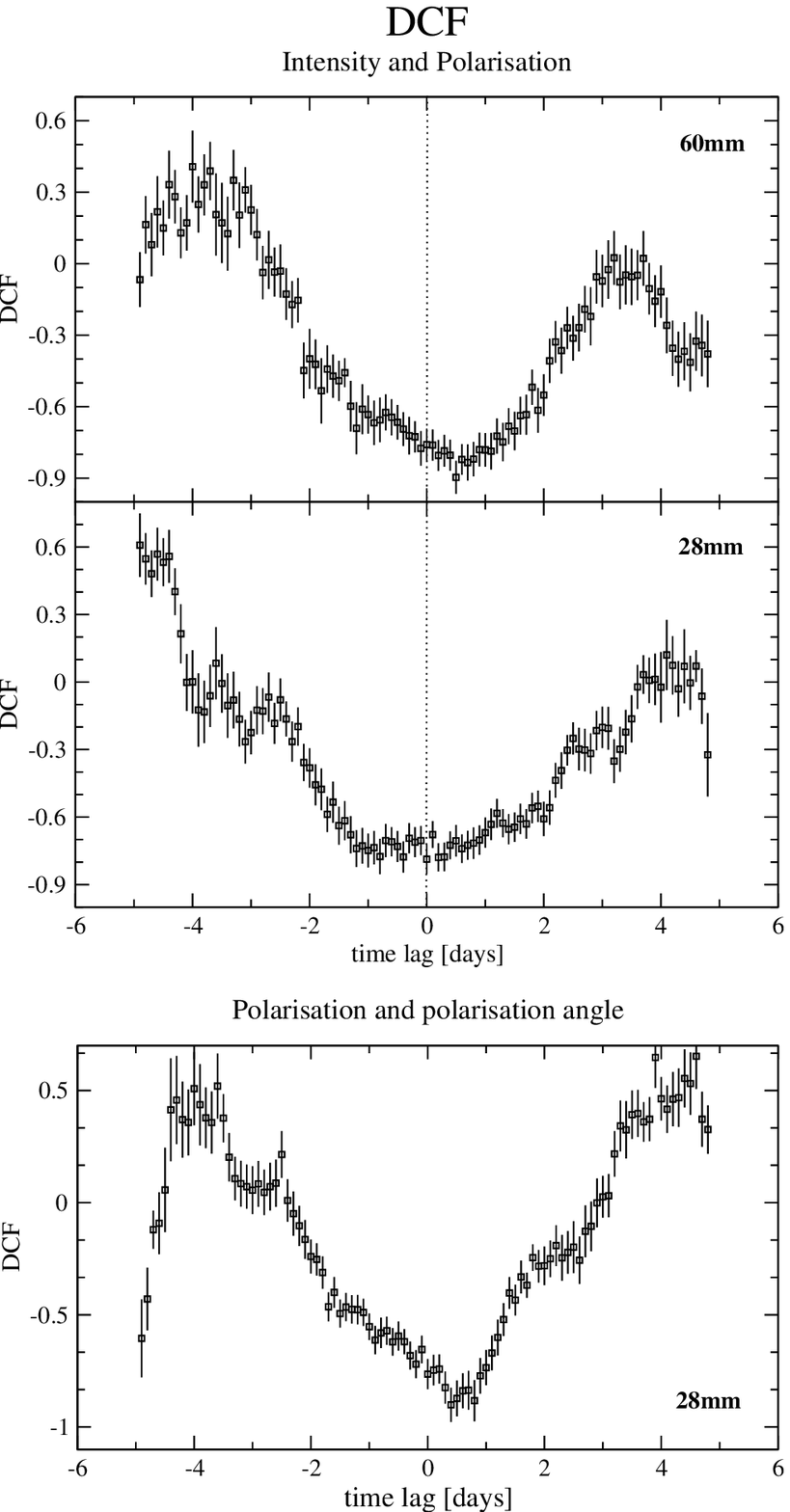}
      \caption{Cross-correlation functions between total intensity and
        polarization at 60\,mm and 28\,mm wavelengths. The top panels show the DCFs
        between total and polarized flux density. The bottom panel shows the
        DCF for polarized intensity and polarization angle at $\lambda$\,28\,mm.}
      \label{ccfs2}
   \end{figure}

In order to investigate the possible anti-correlation of the total intensity
and polarization variations mentioned earlier, we performed a cross-correlation 
analysis (DCF) also between total and polarized intensity, and between polarized 
intensity and polarization angle for the 60 and 28\,mm wavelengths data. The
results are shown in Fig. \ref{ccfs2}. For both radio bands we confirm an
anti-correlation between the total and polarized intensity with a trend of 
the total intensity leading the polarized intensity by 0.5\,days at 60\,mm but no 
obvious time lag at 28\,mm. Formally, we also find an anti-correlation between 
the polarized intensity and polarization angle at 28\,mm wavelength. 

\subsection{Broad-band spectra}\label{spectra}
\begin{figure*}
   \centering
   \vspace{0.5cm}
   \includegraphics[width=8cm,angle=-90]{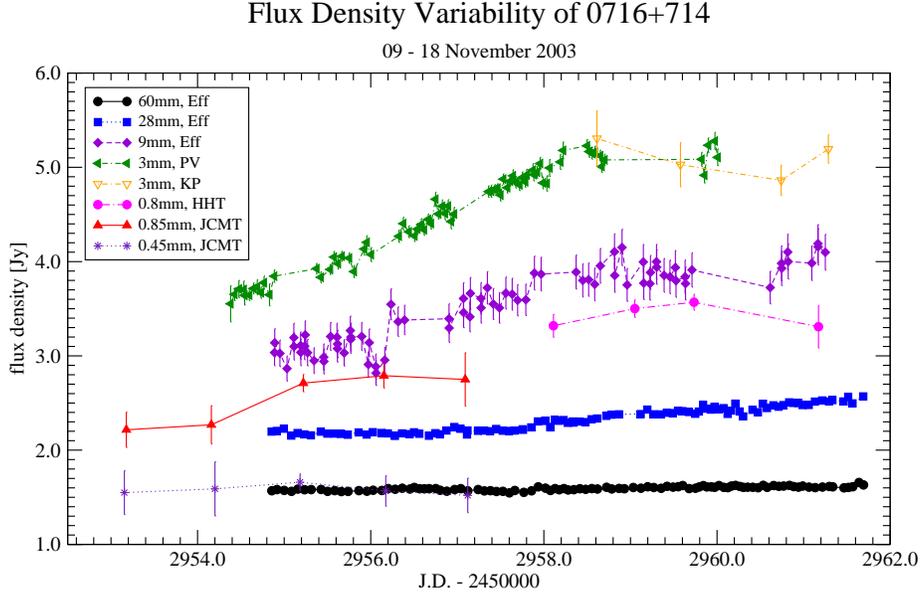}
      \caption{Summary of the total flux density variations in 0716+714 obtained at the various cm- 
	to sub-mm bands during the time of the core campaign. The contribution from each observatory 
	is indicated in the top right insert and in Table \ref{observatories} (see text for details). 
	For better illustration, the light curve at $\lambda$\,1.3\,mm (see Fig. \ref{Fig_LC_1mm}), 
	which is affected by larger measurement errors, is not shown.}
      \label{Comb_LC}
   \end{figure*}

In Fig. \ref{Comb_LC}, we show the combined variability data obtained for 0716+714.
This multi-frequency data set allows to study the spectral evolution and variability
on a daily basis. We construct (quasi-) simultaneous radio spectra
using daily averages from the data of {\it all} participating radio observatories: 
Effelsberg, Pico Veleta, SMT/HHT, JCMT, KP and WSRT (see Table 1). The spectral coverage 
ranges from 1.4\,GHz to 666\,GHz. At 3\,mm wavelength we combined the data from 
IRAM and KP, to extend the time coverage up to November 17. At 0.8 and 0.85\,mm, we 
used the data from JCMT and the SMT/HHT, which unfortunately do not overlap in time.
The time evolution of the cm- to sub-mm spectrum of 0716+714 over the seven 
observing days (November 11--17; hereafter referred to as days 1 to 7) is presented in Fig. 
\ref{specs}. Due to the different duration of the observations at the different telescopes, 
the maximum frequency coverage was obtained only on day 1 (1.4--666\,GHz), whereas for
days 6 and 7 the frequency coverage is reduced to 4.85--345\,GHz. 

0716+714 shows a very inverted radio spectrum over the whole observing period,
with small, but significant brightness variations occurring near the spectral turnover during 
the first five observing days. A pronounced spectral maximum is seen near 90\,GHz. 
Linear fits to the daily spectra yield an averaged spectral slope $\bar{\alpha}_{thick}$ 
($S_{\nu}\sim\nu^{\alpha}$) of $+\,0.39\pm0.02$ for the optically thick part of the spectrum
between 1.4 and 86\,GHz. \citet{2006A&A...456..117A} give for the spectral slope 
between 86\,GHz and 229\,GHz a typical $\bar{\alpha}_{86/229}=-0.23\pm0.10$,
a characteristic for the transition towards optical thin synchrotron emission.
With the broader frequency coverage shown 
in Fig. \ref{specs}, we obtain an average value of $\bar{\alpha}_{thin}=-\,0.30\pm0.05$. 
We note that a more detailed analysis of the high frequency data from JCMT
indicates that the calibration of the 666\,GHz data is uncertain and the
measured fluxes come out very low. For this reason we did not include the 666\,GHz data 
in the spectral fitting.  
%

The spectral slopes obtained from the daily fits suggest slight 
changes with a moderate spectral hardening in the optically thick part
(from $\alpha_{thick}=+0.37$ to $\alpha_{thick}=+0.41$ between days 3 and 4)
and a spectral steepening in the optical thin part
(from $\alpha_{thin}=-0.25$ to $\alpha_{thin}=-0.38$ between days 1 and 3).
For the turn-over frequency $\nu_{m}$ we made parabolic fits to the daily spectra,
which suggest a small shift of $\nu_{m}$ of about 10\,GHz towards lower frequencies.  
However, these changes in $\alpha$
and $\nu_{m}$ appear not to be statistically significant within our measurement 
accuracy and the change in frequency coverage of our spectra. A more detailed 
discussion of the spectrum and its variability will be presented in Sect. 
\ref{spec_bfield}.  
%
%

\section{Discussion}\label{discussion}

\subsection{Variability characteristics}\label{discussion1}

\subsubsection{Variability in the cm- to mm-regime}
During this campaign 0716+714 exhibits very different variability characteristics 
when compared to previous IDV observations in the cm-regime 
\citep[e.g.][]{1992A&A...258..279Q,2003A&A...401..161K}. Usually the source 
shows rapid IDV on time scales of $\lesssim$\,0.5--1.5\,days (in intensity and polarization).
In this observation, however, it showed slow inter-day variability on a $\gtrsim$\,3-4\,day 
time scale (total intensity). When 0716+714 exhibits rapid IDV, the variability 
amplitudes often decreases between 5 and 10\,GHz 
\citep[see Fig. 1 of][]{2002PASA...19...14K}. A variability amplitude constant 
or increasing with frequency was observed only occasionally. It is therefore 
remarkable that this new observations show such a systematic and strong 
increase of the variability amplitudes (from $Y\!=\!3.7$\,\% at 5\,GHz 
to $Y\!=\!34$\,\% at 90\,GHz).

\begin{figure}
   \centering
   \includegraphics[width=9cm,angle=0]{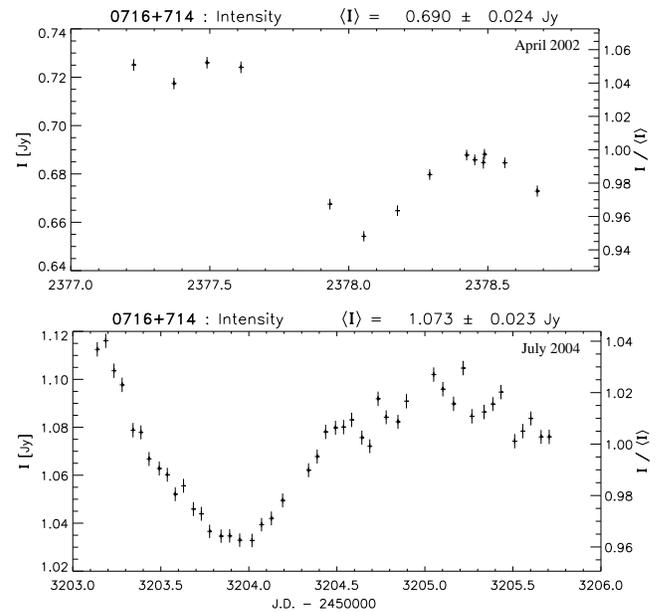}
      \caption{IDV light curves of 0716+714 observed with Effelsberg in April 2002
        (top) and July 2004 (bottom) at $\lambda$\,60\,mm. Note the different and more typical
        (`classical') variability pattern, with faster ($\sim$\,0.5--1\,day) and 
        more pronounced $m=2.1\!-\!3.4$\,\%) variations than those observed in
        November 2003 (Fig. \ref{Fig_LC_6cm}). In the cases shown here, the source was in a  
        fainter radio state than in November 2003, with a mean total flux 
        density of 0.69 and 1.07\,Jy in April 2002 and July 2004, respectively.}  
      \label{6cm_2002_04}
   \end{figure}

A direct comparison with $\lambda$\,60\,mm Effelsberg IDV light curves
obtained earlier (April 2002) and 8 month later (July 2004) reveal the 
`classical' IDV behavior, with variability time scales of 
0.5--1\,days and modulation indices of 2.1--3.4\,\%. These data
sets are shown in Fig. \ref{6cm_2002_04} for comparison. This suggests possible changes 
of the variability characteristics over time scales of weeks to 
months. We note that a change of the variability time scale from $\sim\,1$ to 
$\sim\,7$\,days was also observed in 0716+714 in 1990 
\citep[][]{1991ApJ...372L..71Q,1996AJ....111.2187W}. Other prominent examples of 
changes of the variability mode or episodic IDV are PKS\,0405$-$385 
\citep[][]{2006MNRAS.369..449K} and 0917+624 
\citep[][Fuhrmann et al., in prep.]{1999A&A...352L.107K,2002PASA...19...64F}.

The good frequency coverage of the observations presented here provides  
for the first time a possibility to study with dense time sampling the 
short-term variations from the cm- up to the mm-regime. Here, the correlated 
variability across all bands, the observed frequency dependence of the variability 
amplitudes, and the observed time lag with variations at higher frequencies
appearing earlier (Sect. \ref{dcfs}, Fig. \ref{ccfs}),
argue in favor of a source-intrinsic origin of the observed variations. Such 
`canonical' variability behavior is usually observed in AGN and other compact
radio sources over longer time intervals (weeks to years) and is commonly 
explained by synchrotron-cooling and adiabatic expansion of a flaring component 
or a shock \citep[e.g.][]{1966Natur.211.1131V,1968ARA&A...6..417K,1985ApJ...298..114M}. 
In this framework, \citet{2003A&A...402..151R} found similar variability 
characteristics for the long-term, multi-frequency radio light curves of 0716+714.     
A possible contribution of scintillation effects and thus extrinsic origin 
will be discussed in Sect. \ref{ISS_mod}. A comparison with the simultaneous 
optical R-band data presented by \citet{2006A&A...451..797O} will be given in Sect. 
\ref{broad_band_longterm}.   

The variability brightness temperatures derived for the total intensity data
sets (Table \ref{timescales}, Sect. \ref{Tb-calc}) reveal lower limits of 
$4-8\cdot10^{16}$\,K for the faster variability component observed at 60\,mm 
wavelength, and $10^{13.8}-10^{15}$\,K for the correlated cm-/mm-wavelength
flux density increase on inter-day time scales. We notice a systematic decrease of
$T_{B}$ towards higher frequencies. Since $T_{B}$ was derived using the light 
travel time argument, which implies {\it upper} limits to the size of the region of
variable emission, it remains uncertain if the observed
frequency dependence of the {\it lower} limits of $T_{B}$ also reflects a 
similar frequency dependence of the {\it source-intrinsic} brightness temperature.
The apparent brightness temperatures, however, exceed the IC limit of $\sim$\,10$^{12}$\,K 
\citep[][]{1969ApJ...155L..71K} by at least $\sim 2$ orders of magnitude,
even at mm-wavelengths. If the excessive $T_{B}$ is the result of
relativistic aberration, the source radiation must be strongly Doppler-boosted 
in all of the observed wavebands. The resulting Doppler-factors will be 
discussed in Sect. \ref{doppler-factors}.

From the analysis presented in Sect. 3, we deduce a complex behavior
of the polarization variability. We find (i) more pronounced variability 
amplitudes in polarization than in total intensity (a factor of $\sim$\,15 
at $\lambda$\,60\,mm), (ii) more pronounced polarization variability (in P 
and $\chi$) at $\lambda$\,60\,mm than at $\lambda$\,28\,mm, (iii)
significantly faster variability at $\lambda$\,28\,mm, and (iv) a clear 
anti-correlation between total and polarized flux density at both observing 
bands. Such a complex variability behavior in polarization is often observed 
in IDV sources \citep[e.g.][]{2003A&A...401..161K}, and is interpreted by
a multi-component sub-structure of the emitting region(s). The 
superposition of individually varying and polarized 
sub-components (characterized by their misaligned polarization vectors) could 
in principle reproduce the observed polarization variations. At present 
it is unclear, whether the necessary variability of the individual
sub-components is source-intrinsic \citep[e.g.][]{1993AcASn..33..298Q}, 
solely caused by interstellar scintillation 
\citep[e.g.][]{1995A&A...293..479R,2002ApJ...581..103R}, or results from a mixture 
of these effects \citep[][]{2002ChJAA...2..325Q}. The recent detection of 
polarization IDV in the VLBI core of 0716+714 clearly places the origin of IDV 
in the unresolved VLBI core-region, and strongly supports the idea of multiple 
embedded sub-components of $< 100$\,$\mu$as size \citep[see][]{2006A&A...452...83B}.
It is unclear, however, whether the different time lags seen between the total 
and polarized intensity at $\lambda$\,60\,mm and $\lambda$\,28\,mm have a 
physical meaning. Source-intrinsic opacity effects and a possible spatial 
displacement of the polarized sub-components relative to the total intensity 
component, as often observed for individual jet components with
polarization VLBI \citep[e.g.][and references therein]{2006A&A...452...83B}, 
might provide a simple explanation.
%
%

\subsubsection{Long-term variability and broad band characteristics}\label{broad_band_longterm}
\begin{figure}
   \centering
   \vspace{0.4cm}
   \includegraphics[width=9cm,angle=0]{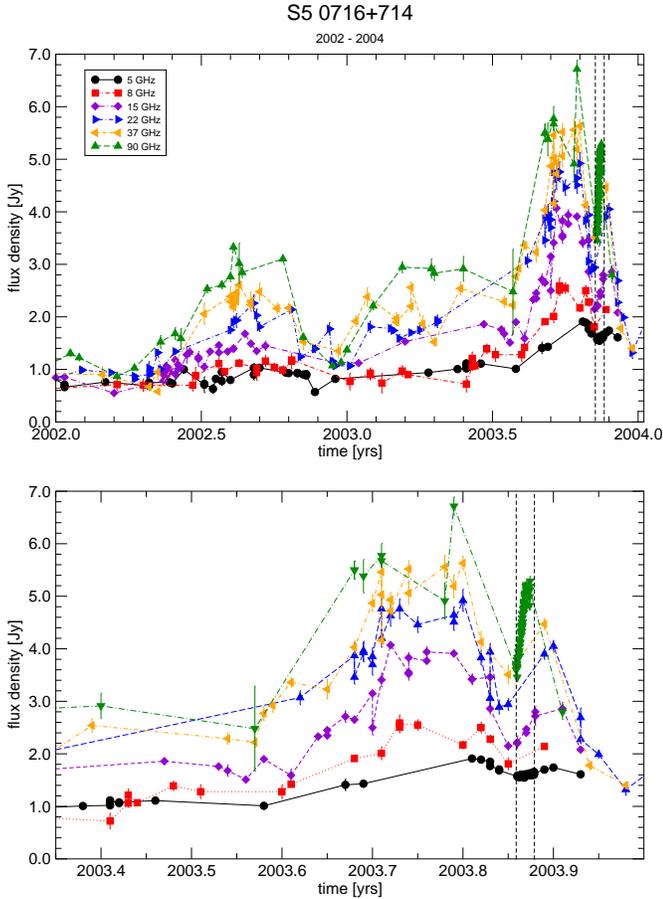}
      \caption{Long-term variability of 0716+714 at 5, 8, 15, 22, 37 and 90\,GHz, showing 
        flux densities observed between 2001.5 and 2004 with the Effelsberg, Mets\"ahovi, Michigan 
        and Pico Veleta radio telescopes. The data sets were taken from Ter\"asranta et al. 2005, 
        Aller \& Aller (private comm.), H. Ungerechts (private comm.), and this paper. Note the unusual 
        source brightness during the outburst around 2003.75. The time of the core campaign 
        (November 8-19, 2003) is marked by dashed vertical lines. For better illustration, an enlargement 
	of the variability curves for the time range 2003.45 -- 2003.95 is shown in the bottom panel.}
         \label{long-term}
   \end{figure}

During the variability campaign in November 2003, 0716+714 was observed in a 
particular active phase, with a high radio-to-optical state. In fact, between 
September and October 2003 the source underwent a dramatic and unprecedented outburst 
in the cm- to mm-bands. In Fig. \ref{long-term} we show the long-term light curves 
obtained between 2001 and 2004 at frequencies ranging between 5 and 90\,GHz. In addition 
to the data from this paper, the figure also includes data from the Michigan monitoring 
(Aller \& Aller, private comm.), the Mets\"ahovi blazar monitoring \citep[][]{2005A&A...440..409T} 
and from IRAM (H. Ungerechts, private comm.). Albeit being in the declining phase of 
the previous large outburst (peak in 2003.8), 0716+714's flux was rising a 
second time and was particularly bright in November 2003, reaching more than 5\,Jy at 
3\,mm wavelength. The overall inverted radio-to-mm spectrum during this time indicates 
a relatively high opacity of the source (synchrotron self-absorption), which is a 
possible cause for the absence of the usually more pronounced IDV in 0716+714 at the 
longer cm-wavelengths.

On VLBI scales 0716+714 frequently shows component ejections on time scales of 
$\sim$\,1--2\,years which often are preceeded by larger flux-density outbursts 
\citep[e.g.][]{2005A&A...433..815B}. Our daily VLBA observations performed during 
November 11--16, 2003 (six epochs), however, indicate no strong structural changes 
at 22 and 43\,GHz and on the mas-scale. The flux density of the VLBI core 
(a $\sim$\,35\,\% increase from day 1 to 6 at 43\,GHz) seems to follow the trend 
which is seen in the total intensity light curves. A more detailed study of the VLBI 
results will be presented in a future paper (Agudo et al., in prep.).

During the campaign, 0716+714 was recorded in a moderate level of optical emission in agreement 
with previous findings, where no obvious strong correlation between prominent radio and optical 
outbursts have been observed in this source \citep[][]{2003A&A...402..151R}. In the optical 
R-band, 0716+714 showed intra-night to inter-day variations. The root mean square variability 
amplitude during the time of the core campaign is $\sim 23$\,\% 
\citep[see Fig. 1 in][]{2006A&A...451..797O}. No obvious correlation between the optical and 
radio variability is seen, indicating that either the emission regions generating radio and 
optical variability are spatially separated, or the variability in both bands has a different 
physical origin. We note that also no obvious simultaneous (or delayed) transition of the 
variability mode is seen between radio and optical, similar to the one reported by \citet{1991ApJ...372L..71Q}.
It is remarkable that during this latter campaign, the source was in a much fainter radio-to-optical 
state and the radio spectrum (between 5 and 8\,GHz) was much less inverted 
\citep[][]{1990A&A...235L...1W,1996ChA&A..20...15Q}. The presence (or absence) of radio-optical 
correlations may depend on the source opacity, which is directly related to the observed shape 
of the radio spectrum and the state of source activity.

Furthermore, we note that in November 2003, 0716+714 was not detected by the INTEGRAL satellite 
at any of its high energy bands (3-200 keV) and only upper limits to the hard X-ray/$\gamma-$ray 
emission could be obtained \citep[see Section \ref{IC_doppler_factors} and][]{2006A&A...451..797O}. 
However 0716+714 was seen by INTEGRAL a few months later (at 30-60\,keV) \citep[April 2004,][]{2005A&A...429..427P},
and some flux density variability in the X-ray band was noted.
%
%

\subsection{Variability versus frequency: comparison with weak interstellar scintillation}\label{ISS_mod}
 
Local scattering in the ISM was clearly demonstrated to be the origin of the most
rapid (sub-hour) variability seen in a few IDV sources \citep[e.g.][]{2001Ap&SS.278..101D}.
Further, ISS on longer time scales (days to weeks) is present in a large number
of radio sources \citep[e.g.][]{2006ApJS..165..439R}. Here, we investigate 
a possible contribution of ISS to the observed cm- up to mm-variability of 0716+714.

\begin{figure}
   \centering
   \includegraphics[width=5.7cm,angle=-90]{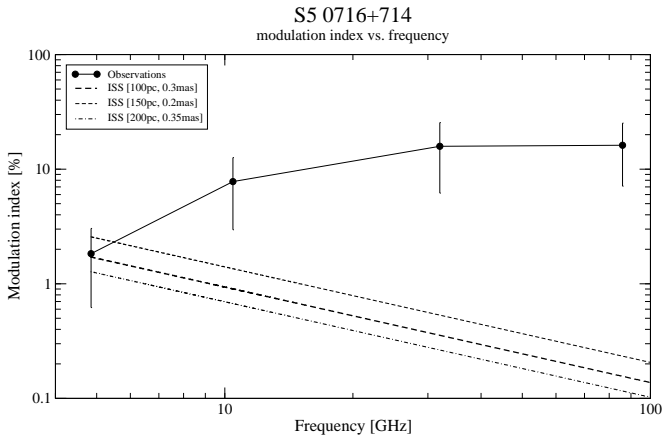}
      \caption{Variability amplitudes of 0716+714 across the observing bands: observed modulation 
        indices versus frequency with ISS model predictions for different parameter ranges (screen distance 
	and source size at $\lambda$\,60\,mm) superimposed. Note the strong discrepancy between model 
	calculations and observed values (see text). The statistical errors of $m$ were obtained taking
        into account the limited total observing time and sampling interval (see Gab\'anyi et al. 
	2007).}
        \label{ISS}
   \end{figure}

In November 2003, 0716+714 varied in a different manner than in previous IDV 
observations: (i) The observed variability index $m$ at $\lambda$\,60\,mm (1.3\,\%) 
is (by a factor $\sim$\,2--4) significantly lower than in most earlier experiments 
\citep[e.g.][]{1992A&A...258..279Q,2003A&A...401..161K,2004PhDT........38K,
2005A&A...433..815B}. Typically, the variability index peaks around 5\,GHz and decreases
towards both higher and lower frequencies. This is expected for the transition between 
weak and strong scattering. The strength of the variability observed in November 2003, 
however, strongly increases with observing frequency (see Sect. \ref{analysis} and \ref{discussion}, 
and Fig. \ref{ISS}). (ii) 0716+714 usually varies fast, on time scales a factor of 
$\sim$\,2--10 faster than observed here. The prolongation of the variability time 
scale in November 2003 might be explained as a seasonal effect, caused by the 
annually changing alignment of the velocity vectors between Earth and scattering screen 
\citep[as observed for e.g. J1819+3845;][]{2002Natur.415...57D}. Screen velocities of 
about 25\,km\,s$^{-1}$ w.r.t. the local standard of rest would be sufficient to explain 
a prolongation in November. However, the multi-frequency time lag (correlation) of the 
variability pattern, the increase of the modulation index from cm- to mm-bands and the low 
variability amplitude at $\lambda$\,6\,cm are not in agreement with such a scenario. As purely 
geometric effect, the orbital motion of the Earth's can affect the variability time scale, 
but not the strength of the observed variations. We also note that 0716+714 is observed 
every 4-8 weeks in a regular IDV monitoring project performed with the Urumqi telescope. 
So far, no strong evidence for annual modulation of the variability pattern is seen 
\citep[][]{2008arXiv0804.2787M}.

At frequencies of a few GHz, the angular size $\theta$ of an extragalactic self-absorbed, 
incoherent synchrotron source is usually expected to be larger than the Fresnel scale $\theta_{F}$ 
for a scattering screen beyond the Solar system \citep[e.g.][]{2002evn..conf...79B}.  
Consequently, we will use a slab-model for weak ISS assuming quenched scattering as described 
by \citet{2002evn..conf...79B}. In the regime of weak, quenched scattering ($\theta>\theta_{F}$), 
the authors give the following analytical solution for the modulation index $m$:
\begin{eqnarray}
m^{2}(\lambda)=2\left(\frac{r_{e}}{D\theta^{2}}\right)^{2}\lambda^{4}
(D\theta)^{\beta-2}SM\cdot f_{1}(\beta)\,\,,
\label{m1}
\end{eqnarray}
with the distance to the scattering screen D, a circular source model with size $\theta$, 
the scattering measure SM, the power law index $\beta=11/3$ for a Kolmogorov spectrum of 
density irregularities and the function $f_{1}(\beta)$ of order unity. Depending on the 
frequency dependence of the source size $\theta$ and for high galactic latitude sources 
(as 0716+714), the amplitude of the ISS induced variations should thus strongly 
decrease towards higher frequencies. 

Assuming a slab of 1\,pc thickness for a thin screen with a strength of turbulence 
$C_{n}^{2}$, similar to the values derived from pulsar scintillation for enhanced scattering 
in the Local Bubble \citep[][]{1998ApJ...500..262B}, we qualitatively compare our multi-frequency 
results of $m$ directly with ISS according to Eq. (\ref{m1}). Combined Space-VLBI 
and single-dish observations of 0716+714 place the physical origin of IDV inside the VLBI-core region
\citet{2005A&A...433..815B}, which contains about 70\,\% of the source's total flux density.
Thus, we assume a scintillating component containing 70\,\% of the observed total flux density, 
and re-scale the observed modulation indices given in Table \ref{results_I_table} according to 
$m_{70\%}=m/0.7$. In this way, we obtain $m_{70\%}=1.8$\,\% at $\lambda$\,60\,mm, which can be 
reasonably reproduced by assuming a scintillating source size of 0.25--0.3\,mas and and a screen 
distance of 100--200\,pc. Taking these values and assuming a linear dependence of $\theta$ with 
frequency, we can compare the observed modulation indices at higher frequencies with those expected 
for weak ISS. The results are shown in Fig. \ref{ISS}. A comparison between observations and model 
strongly suggests that the ISS model can not reproduce the increasing variability amplitudes observed 
towards higher frequencies. We obtain similar results by systematically changing the model parameters 
such as screen distance, source size and scattering measure. Consequently, we conclude that the 
observed simultaneous and correlated inter-day variations of 0716+714 on time scales of several 
days in the cm- to mm-band are not strongly affected by scintillation effects, and therefore 
should be considered as intrinsic to the source.  

If ISS dominates the variability at $\lambda$\,60\,mm, however, the low variability amplitude 
and the long variability time scales (compared to previous observations) must indicate strong 
changes, either in the scattering medium, or in the intrinsic source structure on scales 
comparable to the scattering size of the medium. Changes in the scattering properties of 
the screen can be related to either changes in the distance or the strength of turbulence. 
Both appear unlikely to happen on timescales as short as months. Since 0716+714 was 
in a flaring state during this observations (see Sect. \ref{discussion1}), changes in the 
source structure therefore appear more likely. Flux density outbursts are often accompanied by 
the ejection of new jet components which, for instance, would temporarily increase the size of the 
scintillating core region. In such a scenario, the size of the scintillating component should have 
increased by a factor 2--3 (from about 0.08--0.09\,mas to 0.25--0.3\,mas) in order to quench $m$, 
down to the observed level of 1.8\,\%. The observations performed eight months later 
(July 2004, Fig. \ref{6cm_2002_04}) reveal a higher modulation index of $m_{70\%}=3.0$\,\%, 
indicating again a smaller scintillating source size of about 0.17--0.19\,mas (assuming a screen 
distance of 100--200\,pc). Future VLBI studies with highest possible angular resolution will 
be needed to relate component ejection, motion and core size evolution with the changes of the 
IDV pattern in 0716+714.
%
%

\subsection{Synchrotron spectra and spectral evolution} \label{spec_bfield}
\begin{figure}
   \centering
   \vspace{0.2cm}
   \includegraphics[width=8.7cm,angle=0]{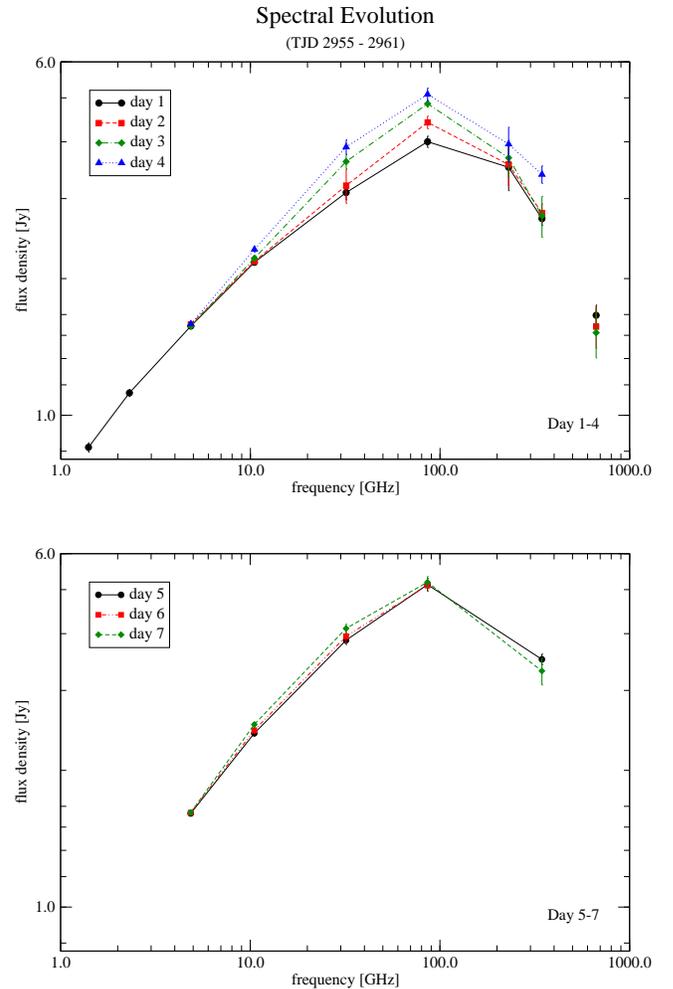}
      \caption{Simultaneous cm- to sub-mm spectra of 0716+714 derived from
        daily averages of the flux-density measurements shown in Fig. \ref{Comb_LC}. 
        The top-panel shows the spectra for days 1--4 (November 11--14,
        J.D.\,$=$\,2452955-58), whereas the spectra of days 5--7 (November 15--17, 
        J.D.\,$=$\,2452959-61) are shown below. The flux density errors are small
        and when invisible, comparable or smaller than the size of the symbols.}
      \label{specs}
   \end{figure}

In Fig. \ref{specs} we show the evolution of the cm- to sub-mm radio spectra
on a daily basis. All spectra are strongly inverted, peaking at $4-5$\,Jy 
near 3\,mm wavelength (90\,GHz). The results of the spectral fits described 
in Sect. \ref{spectra} demonstrate that during the whole core campaign the radio 
spectrum was optically thick up to the mm-band with $\bar{\alpha}_{thick}=+\,0.39\pm0.02$ 
and optically thin ($\bar{\alpha}_{thin}=-\,0.30\pm0.05$) beyond a synchrotron 
turnover frequency $\nu_{m}$ somewhere close to 90\,GHz. This confirms previous 
findings by \citet{2006A&A...451..797O}, who showed the broad band spectral energy 
distribution (from radio to $\gamma$-rays) using the data of this campaign. The radio 
spectra shown here (Fig. \ref{specs}) largely differ from earlier multi-frequency 
studies, where the source typically showed a flat and only mildly inverted (1\,Jy) 
spectrum \citep[e.g.][]{1988A&A...192L...1C,1990A&A...235L...1W,1991ApJ...372L..71Q}.
In contrast, the new spectra are indicative of a high peaking and self-absorbed
flare spectrum.
%
%

Significant changes in the cm- to sub-mm spectra occur mainly until day 5, 
i.e. during the period of the monotonic increase in total flux density seen 
across all bands. Here, the spectral peak flux increases,
reaching its maximum on day 5. The daily averaged peak flux density (at $\nu_{m}$) rises 
from $S_{86\,GHz}=4.0$\,Jy on day 1 to $S_{86\,GHz}=5.1$\,Jy on day 5. 
After this date, no further changes are evident (see also Fig. \ref{Comb_LC}). 

The absence of pronounced variations in the spectral slopes and the turnover frequency 
(achromatic variations) would indicate a geometrical origin, i.e. caused by changes in 
the beaming factor \citep[e.g. in a helical/precessing jet;][]{1980Natur.287..307B,1999A&A...347...30V}. 
The possible slight changes in the daily spectral slopes (spectral steepening for $\nu\,<\,\nu_{m}$, 
the spectral softening for $\nu\,>\,\nu_{m}$) and marginal shift of the turn-over frequency 
towards lower energies ($\Delta \nu \sim 10$\,GHz), suggest that the spectral evolution is 
not fully achromatic. The limited time coverage (only 7 days) prevents a more detailed analysis 
of the spectral evolution. However, the presence of source intrinsic spectral variations is 
also supported by the canonical behavior of the time lag (discussed in Sect. \ref{dcfs}) and 
the related frequency dependence of the variability amplitudes, which increase towards the mm-bands. 
Therefore, an interpretation within standard flare/adiabatic expansion and/or shock-in-jet models 
appears applicable \citep[e.g.][]{1985ApJ...298..114M}.

The optically thick spectral index, however, strongly differs from the canonical value for a 
self-absorbed, homogeneous synchrotron component ($S_{\nu}\sim\nu^{5/2}$). 
This indicates that the flaring component, which dominates the observed radio-spectrum, 
is inhomogeneous and may consists of more homogeneous but smaller sub-components, peaking at 
slightly different self-absorption frequencies \citep[see also][]{2006A&A...451..797O}.
This view is further supported by (non-simultaneous) mm-VLBI observations of 0716+714, 
which indeed show multiple components on the 0.05 - 0.1\,mas scale 
(T. Krichbaum, unpublished data).

\subsection{Doppler factors and magnetic field}\label{doppler-factors}

In this section we will compare the Doppler factors $\delta_{var}$ deduced 
from the variability brightness temperatures of Sect. \ref{Tb-calc} with Doppler 
factors calculated by other methods. In particular, we will compare $\delta_{var}$ 
with (i) estimates derived from VLBI studies of 0716+714 ($\delta_{VLBI}$), (ii) the 
equipartition Doppler factor $\delta_{eq}$, using calculations of the synchrotron 
and equipartition magnetic field, (iii) $\delta_{var,eq}$ using the equipartition brightness 
temperature $T_{eq}$, and (iv) the inverse Compton Doppler factor $\delta_{IC}$ as calculated 
from the simultaneous INTEGRAL observations. 

\subsubsection{Doppler factor $\delta_{var,IC}$}\label{T_B_doppler}
The limits of the apparent brightness temperature ($T_{B}^{app}$)  summarized in 
Table \ref{timescales} exceed by several orders of magnitude the IC-limit of 
$T_{B,IC}^{lim}\sim$\,10$^{12}$\,K \citep[][]{1969ApJ...155L..71K}. 
In order to explain the excessive temperatures solely by relativistic boosting 
of the radiation, source intrinsic radiation near the IC-limit can be assumed 
($T_{B}=T_{B,IC}^{lim}$), which leads to a lower limit to the Doppler factor of 
the emitting region of $\delta_{var,IC}=(1+z)\sqrt[3+\alpha]{T_{B}^{app}/10^{12}}$. 
This leads to Doppler factors $\delta_{var,cm}>$\,10--22 in the cm-regime and 
$\delta_{var,mm}>$\,5--10 in the mm-regime (assuming z\,$=$\,0.3), which lower the 
observed brightness temperatures of Table \ref{timescales} down to the IC-limit. In the 
case of the faster variability component seen at $\lambda\,60$\,mm, the situation becomes 
more extreme. Here, Doppler factors $\delta_{var,IC}$ $>$\,43 would be needed if, again, 
sole intrinsic in origin. Although high, these limits for $\delta_{var,cm}$ and 
$\delta_{var,mm}$ agree well with recent estimates coming from kinematic VLBI studies 
of the source. From 26 epochs of VLBI data observed between 1992 and 2001, 
\citet{2005A&A...433..815B} found atypically high superluminal motion 
and conclude that the minimum Doppler factor is $\delta_{VLBI}\geq $\,20--30,
assuming a redshift of $z \geq 0.3$. This confirms the previously reported high 
speeds (16--21\,c) in 0716+714 \citep[][]{2001ApJS..134..181J}. Moreover, using a 
larger sample of sources, \citet{2005AJ....130.1418J} observed $\delta_{VLBI}>$\,16--21 
in 9 of the 13 studied blazar indicating that strong Doppler boosting is not a rare 
phenomenon in compact VLBI jets. 

We further note that the observed decrease of $T_{B}^{app}$ towards higher
frequencies suggests that the Doppler factors at cm- and mm-wavelengths could be 
different. This appears not unlikely in a model of a stratified
and optically thick self-absorbed bent jet, where at each wavelength 
spatially different jet regions are probed. Each of these regions may exhibit
characteristic Doppler factors depending on jet speed and viewing angle. 
As one looks at higher frequencies into regions which are deeper embedded in
the source, a lower brightness temperature and resulting lower $\delta_{var}$ 
towards the innermost jet regions would then indicate either jet-acceleration
(changes of the Lorenz factor along the jet) and/or jet bending with changes
of the jet orientation away from the observer`s line of sight for outward
oriented motion.     

\subsubsection{Magnetic field from synchrotron self-absorption}

Using standard synchrotron expressions, it is possible to constrain the magnetic
field B in 0716+714 by means of the spectral parameters deduced in the previous section. 
Assuming a homogenous synchrotron self-absorbed region, a closed expression for
B is given by \citep[see e.g.][]{1987slrs.work..280M}
\begin{eqnarray}
B_{SA}\,[G]\,=10^{-5}b(\alpha)\theta^{4}\nu_{m}^{5}S_{m}^{-2}\left(\frac{\delta}{1+z}\right),
\label{B}
\end{eqnarray}
with the tabulated quantity b$(\alpha)$ depending on the optically thin
spectral index $\alpha_{thin}$ \citep[see Table 1 in][]{1987slrs.work..280M}, 
the flux density $S_{m}$, the source angular size $\theta$ at the synchrotron turnover 
frequency $\nu_{m}$ and the Doppler factor $\delta$. At $\nu_{m}=86$\,GHz, 
the size of the emitting region responsible for the observed variations 
can be constrained using recent mm-VLBI measurements of the core region of
0716+714 \citep[][]{2006A&A...452...83B,2006A&A...456..117A} which yield 
$\theta_{86\,GHz}<0.04$\,mas. Using the observed spectral parameters from 
above ($S_{86\,GHz}\!=\!4.0-5.13$\,Jy, $\bar{\alpha}_{thin}\!=\!-\,0.30$,) we calculate a 
lower limit of the magnetic field $B_{SA}$ in the range of $0.07-0.11\,\delta\,$G for 
$\nu_{m}\!=\!86$\,GHz without correcting for Doppler boosting. 

\subsubsection{Magnetic field from equipartition}
   
The magnetic field $B_{SA}$ calculated according to Eq. (\ref{B}) can be used to 
constrain the Doppler factor in an alternative way. The magnetic equipartition
field $B_{eq}$, which minimizes the total energy 
$E_{tot}=(1+k)E_{e}+E_{B}$ (with relativistic particle energy $E_{e}\sim B^{-1.5}$
and energy of the magnetic field $E_{B}\sim B^{2}$), is given by the following
expression  \citep[e.g.][]{1970ranp.book.....P,2005A&A...433..815B}:
\begin{eqnarray}
B_{eq}&=&\left(4.5\cdot(1+k)\,f(\alpha,\nu_{a},\nu_{b})\,L\,R^{3}\right)^{2/7}\nonumber\\
&=& 1.65\cdot10^{10}\left(f(\alpha,\nu_{a},\nu_{b})\,(1+z)\,S_{m}\,\nu_{m}\,d_{L}^{2}\,R^{-3}\right)^{2/7}
\label{B_min}
 \end{eqnarray}
with the energy ratio $k$ between electrons and heavy particles, the synchrotron luminosity 
$L=4\pi\,d_{L}^{2}(1+z)\int^{\nu_{a}}_{\nu_{b}}{S\,d\nu}$ of the source, the size of the 
component R in cm, $S_{m}$ and $\nu_{m}$ in GHz and the tabulated function 
$f(\alpha,\nu_{a},\nu_{b})$ with the upper and lower synchrotron frequency
cutoffs $\nu_{a},\nu_{b}$. Using $\nu_{a}=10^{7}$\,Hz, $\nu_{b}=5.5\cdot10^{14}$\,Hz 
\citep[e.g.][]{1987ApJ...322..643B,1993ApJ...407...65G}, $k\approx100$ and the 
same values as for Eq. (\ref{B}), we calculate a lower limit for the magnetic field 
in the range of 1.2--1.3\,G, which is higher than the magnetic field obtained in the 
previous section. Eq. (\ref{B}) and (\ref{B_min}) give different dependencies
of $B$  and $\delta$: $B_{SA}\sim\delta$, $B_{eq}\sim\delta^{2/7\alpha+1}$.
This yields  $B_{eq}/B_{SA}=\delta_{eq}^{2/7\alpha}$. Adopting the above numbers,
we obtain Doppler factors $\delta_{eq}$ in the range of 12--23,
in good agreement with both, the previous estimates for $\delta_{var}$ as well
as $\delta_{VLBI}$.

\subsubsection{Equipartition brightness temperature $T_{B,eq}^{lim}$}

Instead of using the IC limit of $10^{12}$\,K for the intrinsic brightness 
temperature, a brightness temperature limit based on equipartition
between particle energy and field energy could be used \citep[][]{1977MNRAS.180..539S,1994ApJ...426...51R}:
$T_{B,eq}\sim\,3\cdot10^{11}$\,K.  It is argued that this limit better reflects
the stationary state of a synchrotron source and for many sources is 
$\lesssim\,10^{11}$\,K \citep[e.g.][]{1994ApJ...426...51R,1996ApJ...461..600G}.
With these numbers we obtain another estimate of the Doppler factor: 
$\delta_{var,eq}=(1+z)\sqrt[3+\alpha]{T_{B}^{app}/3\cdot10^{11}}>$\,8--33, 
where we again used $T_{B}^{app}$ from Table \ref{timescales}. Although high, 
these values still agree with the previous estimates and also with $\delta_{VLBI}$.

\subsubsection{Inverse Compton Doppler factor $\delta_{IC}$}\label{IC_doppler_factors}

Finally we can independently calculate the Doppler factor $\delta_{IC}$ using the upper limits
to the soft $\gamma$-ray flux densities of 0716+714 obtained from the INTEGRAL observations 
during this experiment and the inverse Compton argument \citep[see also][]{2006A&A...451..797O,2006A&A...456..117A}.
Following \citep[see e.g.][]{1987slrs.work..280M,1993ApJ...407...65G} the IC Doppler factor 
$\delta_{IC}$ is calculated from:
\begin{eqnarray}
\delta_{IC}=f(\alpha)S_{m}\left(\frac{ln(\nu_{b}/\nu_{m})\nu_{\gamma}^{\alpha}}{S_{\gamma}\theta_{\nu}^{(6-4\alpha)}\nu_{m}^{(5-3\alpha)}}\right)^{1/(4-2\alpha)}(1+z)\,\,.
\label{doppler_IC}
\end{eqnarray}
where $\nu_{b}$ is the synchrotron high frequency cut-off in GHz, $S_{m}$ the flux 
density in Jy at the synchrotron turnover frequency $\nu_{m}$, $S_{\gamma}$ the observed 
$\gamma$-ray flux in Jy at $\nu_{\gamma}$ in keV, $\theta$ the source angular size in mas 
and a function of spectral index $f(\alpha)\simeq0.14-0.08\alpha$ 
\citep[e.g.][]{1993ApJ...407...65G}. If we determine the size $\theta_\nu$ via
the causality argument and with the variability time scale $t_{var}$, an additional
$\delta$ comes in Eq. (\ref{dim}) \citep[see also][]{2006A&A...456..117A} and we can 
rewrite Eq. (\ref{doppler_IC}):
\begin{eqnarray}
\delta_{IC}=\left[f(\alpha)S_{m}(1+z)\right]^{(4-2\alpha)/(10-6\alpha)}\left(\frac{ln(\nu_{b}/\nu_{m})\nu_{\gamma}^{\alpha}}{S_{\gamma}\theta^{(6-4\alpha)}\nu_{m}^{(5-3\alpha)}}\right)^{1/(10-6\alpha)}\,\,.
\label{doppler_IC_2}
\end{eqnarray}
whereas now the apparent variability size is calculated from 
$\theta\equiv0.13\,t_{var}d_{L}^{-1}\,(1+z)$. Using the same values for 
$\nu_{m}$, $\nu_{b}$, $\alpha_{thin}$ as above and $S_{86\,GHz}\!=\!5.13$\,Jy, 
$t_{var}=10.7$\,days (Table \ref{timescales}), we calculate $\delta_{IC}$ using 
the upper limits to the soft $\gamma$-ray flux in the four bands of INTEGRAL
\citep[8, 23, 63 and 141\,keV, see also][]{2006A&A...451..797O}.
The upper limits to the  $\gamma$-ray flux lead to lower limits to
the IC Doppler-factors, which for the different energy bands are 
summarized in Table \ref{delta_all}. As outlined by \citet{2006A&A...456..117A},
the lower limit of $\delta_{IC}\!>\!14.1$ obtained in this way provides a more
robust constraint on the Doppler factor than $\delta_{var}$ and $\delta_{eq}$
due to the weak dependence of Eq. (\ref{doppler_IC_2}) on z and $d_{L}$ for 
$\alpha_{thin}\!=\!-0.3$. We further note that the values of $\delta_{IC}\!>\!14$ 
obtained here are in very good agreement with a Doppler factor of
$\delta\,\sim\,17$ derived by \citet{2003A&A...400..477T}, who fit the 
SEDs of 0716+714 by a homogenous, one-zone synchrotron/inverse Compton 
model. 

\begin{table}
\begin{center}
\caption{Summary of the Doppler factor estimates using the 
different methods presented in Sect. \ref{doppler-factors}. See text for
details.}
\begin{tabular}{l|ll}
\hline
\hline
$\delta$               & value       & used values/quantities\\
\hline
$\delta_{var,IC,cm}$      & $>$\,10--22 & $T_{B}^{app}$ (Table \ref{timescales}),
$T_{B,IC}^{lim}=$\,10$^{12}$\,K \\      
$\delta_{var,IC,mm}$      & $>$\, 5--10 & $T_{B}^{app}$ (Table \ref{timescales}),
$T_{B,IC}^{lim}=$\,10$^{12}$\,K \\
$\delta_{var,eq}$         & $>$\, 8--33 & $T_{B}^{app}$ (Table \ref{timescales}),
$T_{B,eq}^{lim}=\,3\cdot10^{11}$\,K \\      
\hline
$\delta_{eq}$             & $>$\,12--23 & $B_{SA}$, $B_{eq}$, $S_{86\,GHz}\!=\!4.0-5.13$\,Jy,\\ 
                          &             & $\alpha_{thin}\!=\!-0.3$, $\theta_{VLBI}<0.04$\,mas  \\
\hline
$\delta_{IC,8\,keV}$   & $>$\,15.9   & $S_{8\,keV}=1.37\cdot10^{-7}$\,Jy, $\alpha_{thin}\!=\!-0.3$\\
                       &             & $S_{86\,GHz}\!=\!5.13$\,Jy, $t_{var}=10.7$\,days\\
$\delta_{IC,23\,keV}$  & $>$\,14.7   & $S_{23\,keV}=2.55\cdot10^{-7}$\,Jy, $\alpha_{thin}\!=\!-0.3$ \\
                       &             & $S_{86\,GHz}\!=\!5.13$\,Jy, $t_{var}=10.7$\,days\\
$\delta_{IC,63\,keV}$  & $>$\,15.3   & $S_{63\,keV}=1.23\cdot10^{-7}$\,Jy, $\alpha_{thin}\!=\!-0.3$ \\
                       &             & $S_{86\,GHz}\!=\!5.13$\,Jy, $t_{var}=10.7$\,days\\
$\delta_{IC,141\,keV}$ & $>$\,14.1   & $S_{141\,keV}=2.42\cdot10^{-7}$\,Jy, $\alpha_{thin}\!=\!-0.3$ \\
                       &             & $S_{86\,GHz}\!=\!5.13$\,Jy, $t_{var}=10.7$\,days\\
\hline
\hline
\end{tabular}
\label{delta_all}
\end{center}
\end{table}

In Table \ref{delta_all} we summarize the Doppler-factors $\delta$ calculated using the 
different methods presented above. A comparison of the numbers shows a general good agreement 
with the Doppler factor $\delta_{VLBI}$ derived from the superluminal VLBI kinematics. 
Although strong Doppler boosting is required, these values appear to be not excessively
high, and therefore provide a self-consistent explanation of the high apparent variability 
brightness temperatures, without a violation of the theoretical limits. Since no excessive 
$\gamma-$ray emission and therefore no IC catastrophe was recorded by INTEGRAL, and a strong 
contribution of ISS is unlikely, we thus conclude that the violation of the theoretical 
limits inferred from our observations was only apparent and intrinsic Doppler boosting 
can naturally explain the non-detection of the source at high energies. However, we 
stress that all derived values of $\delta$ must be regarded as lower limits. Consequently, 
Doppler-factors much higher than $\delta>20-30$ can still not be ruled out completely. 
Contributions from other radiation processes, such as partially coherent emission, 
can not be ruled out either.           

\section{Summary and Conclusions}

We presented the results of a combined variability analysis from an intensive 
multi-frequency campaign on the BL Lac 0716+714 performed during seven days in 
November 2003. From seven participating radio observatories we obtain a frequency 
coverage of 1.4--666\,GHz. Densely sampled IntraDay Variability (IDV) light curves
obtained at wavelengths of 60, 28, 9, 3, and 1.3\,mm allowed for the first time a 
detailed analysis of the source's intra- to inter-day variability behavior over the 
full radio- to short mm-band. In this observing campaign 0716+714 was found to be 
in a particular slow mode of variability, when compared to all previous IDV 
observations of this source. While in total intensity a component of faster 
variability was observed only at $\lambda$\,60\,mm, the source's flux density in 
the cm- to mm-regime was dominated by a nearly monotonic increase on inter-day time 
scales and with variability amplitudes strongly increasing towards higher frequencies. 
Here, our CCF analysis confirms that the flux density variations are correlated across the 
observing bands, with variability at shorter wavelengths leading. This and the 
observed frequency dependence, which cannot be explained by a model for weak 
scintillation, strongly suggest that the observed inter-day variability has to be 
considered as source-intrinsic rather than being induced by ISS. Only at 60\,mm 
wavelength a component of faster variability is seen and implies an unusually high 
apparent brightness temperature $T_{B}$. Hence, ISS might be present at this frequency.     

The non-detection of the `classical', more rapid (type II) IDV behavior of
0716+714 in the cm-bands is most likely caused by opacity effects and can be related 
to the overall flaring activity of the source shortly before and during this
campaign. Since episodic IDV behavior is also observed in other sources 
\citep[e.g.][]{2002PASA...19...64F,2006MNRAS.369..449K}, it appears likely 
that the variability pattern in `classical' type\,II IDV sources is strongly affected 
by the evolution of their intrinsic complex structure on time scales of weeks to months. 
The observed complicated variability patterns in total intensity and polarization, and 
their correlations, indicate the existence of a multi-component structure with individually 
varying and polarized sub-components of different size.

From daily averages, the spectral evolution of the highly inverted radio-to-sub-mm spectrum 
of 0716+714 could be studied. During the seven observing days the spectrum always peaked near 
$\nu_{m}\!\sim\!86$\,GHz. Significant changes of the peak flux mainly occurred during 
the first 5 days with a continuous {\bf rise} of the peak flux density from 4 to 5\,Jy. 
Together with possibly small changes in the daily spectral slopes and the peak frequency $\nu_{m}$,
the observed variations follow the `canonical' behavior and indicate time-variable
synchrotron self-absorption and adiabatic expansion of a shock or a flaring component
as described by standard models \citep[e.g.][]{1985ApJ...298..114M}.

The apparent brightness temperatures $T_{B}$ obtained from the inter-day variations 
exceed theoretical limits by several orders of magnitudes. Although $T_{B}$
decreases towards the mm-bands, the $10^{12}$\,K IC-limit is always violated. 
Assuming relativistic boosting of the radiation, the source must always be
strongly Doppler boosted. We obtain lower limits to the Doppler factor of the source
using different methods, including (i) inverse Compton ($\delta_{var,IC}$) and
equipartition ($\delta_{var,eq}$) estimates using the variability brightness temperatures,
(ii) an estimate  $\delta_{eq}$ using calculations of the synchrotron and
equipartition magnetic field, and (iii) an inverse Compton Doppler factor $\delta_{IC}$
using the data from the simultaneous INTEGRAL observations. These methods 
reveal robust and self-consistent lower limits to the Doppler factor 
with $\delta_{var,IC}>$\,5, $\delta_{var,eq}>$\,8 and $\delta_{eq}>$\,12. 
These limits are in good agreement with estimates based on recent kinematical 
VLBI studies of the source and the IC Doppler factor $\delta_{IC}>$\,14 
obtained from the upper limits to the high energy emission in the 3--200\,keV bands.
The non-detection of the source in the soft $\gamma$-ray bands implies the absence of 
a simultaneous strong IC catastrophe during the period of our IDV observations. 
Since a strong contribution of interstellar scintillation to the observed inter-day variability can 
be excluded, we conclude that relativistic Doppler boosting appears to naturally explain 
the observed apparent violation of the theoretical brightness temperature limits.  

\begin{acknowledgements}
      	This work was partly supported by the European Institutes belonging to the ENIGMA 
      	collaboration, who acknowledge EC funding under contract HPRN-CT-2002-00321.
      	This research is based on observations with the 100-m telescope of the MPIfR 
      	(Max-Planck-Institut f\"ur Radioastronomie) at Effelsberg. This research has made 
	use of data from the University of Michigan Radio Astronomy Observatory which 
	has been supported by the University of Michigan and the National Science Foundation. 
	We gratefully thank M.F. Aller and H.D. Aller for providing these UMRAO flux-density
     	data. This work has made use of observations with the IRAM 30-m telescope, the JCMT and
        the telescopes of the Arizona Radio Observatory (SMT/HHT and KP-12m). L.O. gratefully 
	acknowledges partial support from the INFN grant PD51. L.O. acknowledges the 
	hospitality of the Landessternwarte Heidelberg-K\"onigstuhl and Tuorla Observatory, 
	where part of this work was done.

\end{acknowledgements}

\appendix
\section{Statistical variability analysis}\label{Ap1} 
As a criterion for the presence of variability, the hypothesis of a constant 
function (non-variable source) is examined and a $\chi^{2}$-test of the
variability curve $S_{i}$ is done: 
\begin{eqnarray}
\chi^{2}=\sum_{i=1}^{N}\left(\frac{S_{i}-<S>}{\Delta S_{i}}\right)^{2}\,\,,
\label{chi}
\end{eqnarray}
with the reduced $\chi_{r}^{2}$
\begin{eqnarray}
\chi_{r}^{2}=\frac{1}{N-1}\sum_{i=1}^{N}\left(\frac{S_{i}-<S>}{\Delta S_{i}}\right)^{2}\,\,.
\label{chi2}
\end{eqnarray}
Here, $S_{i}$ denotes the individual flux density measurements at time $i$, 
$<S>$ their average in time, $\Delta S_{i}$ the individual measurement errors and N 
the number of measurements \citep[e.g.][]{1982aite.book.....T,1992drea.book.....B}.
A source is considered to be variable if the $\chi^{2}$-test gives a probability 
of $\le 0.1\,\%$ for the assumption of constant flux density (99.9\,\%
significance level for variability).

To further quantify the observed variability, the modulation index $m$ is calculated. 
This quantity provides a useful measure of the strength of the observed variations and 
is defined as the percentage ratio of the rms-fluctuations $\sigma_{S}$ of the time 
series and the time averaged mean flux density $<S>$:
\begin{eqnarray}
m[\%]=100\cdot \frac{\sigma_{S}}{<S>}\,\,.
\label{m}
\end{eqnarray}
In order to take the residual calibration errors into account, we follow 
\citet{1987AJ.....94.1493H} and calculate the variability amplitude $Y$ for each individual 
light curve:
\begin{eqnarray}
Y[\%]=3\sqrt{m^{2}-m_{0}^{2}}\,\,.
\label{y}
\end{eqnarray}
The $Y$-amplitude can be regarded as being equivalent to a noise-bias corrected 
variability amplitude using the mean modulation index $m_{0}$ of the secondary 
calibrators as a measure of the overall calibration accuracy.

A similar analysis was performed for polarization. The corresponding quantities 
for Polaris ed intensity $S_{P}$ are
\begin{eqnarray}
m_{P}[\%]=100\cdot \frac{\sigma_{S_{P}}}{<S_{P}>}\,\,,
\label{m_P}
\end{eqnarray}
\begin{eqnarray}
Y_{P}[\%]=3\sqrt{m_{P}^{2}-m_{P,0}^{2}}\,\,.
\label{y_P}
\end{eqnarray}
For the the polarization angle $\chi$ we use 
\begin{eqnarray}
m_{\chi}[^\circ]=\sigma_{\chi}\,\,,
\label{m_chi}
\end{eqnarray}
\begin{eqnarray}
Y_{\chi}[^\circ]=3\sqrt{\sigma_{\chi}^{2}-\sigma_{\chi,0}^{2}}\,\,.
\label{Y_chi}
\end{eqnarray}
A $\chi^{2}$-test for $S_{P}$ and $\chi$ was performed 
similar to total intensity, i.e. using Eq. (\ref{chi}) and (\ref{chi2}). 
  
\section{Variability time scales}\label{Ap2} 

{\noindent \it (i) Structure function analysis}\\
\\
Following \citet{1987AJ.....94.1493H} and \citet{1999A&A...352L.107K}, 
the time series were analyzed by first order structure functions (SF). 
The following procedure was used and applied to each data set in order 
to obtain a variability time scale at every frequency 
\citep[see also][]{2002evn..conf...79B}. For each data set $f(t_{i})$, the 
discrete structure function
\begin{eqnarray}
SF(\tau_{j})&=&N^{-1}_{ij}\sum_{i=1}^{n}w(i)w(i+j)[f(t_{i})-f(t_{i}+\tau_{j})]^{2}
\label{sf3}
\end{eqnarray}
was calculated. Here, $N_{ij}$ is the normalization, $t_{i}$ and $t_{i}+\tau_{j}$ 
are the times at which the flux densities $f$ were obtained, and $w(i)$ and 
$w(i+j)$ are weighting functions with $w(i)w(i+j)>0$. The weighting function 
accounts for binning of an unevenly sampled time series. The binning 
interval was chosen to be $\sim$\,3--5 times of the typical sampling interval 
of the light curve. In addition, the structure functions were also calculated 
using an interpolation method similar to that described by \citet{1988ApJ...333..646E}.
This showed no significant differences in the results compared to the discrete 
method. Subsequently, two linear fits were performed to estimate the
slope of the SF ($SF=a\cdot \tau^{\beta}$) and the plateau level
($\rho_{0}=2m^{2}$). This plateau (or saturation) level is determined through calculation
of the mean value of the SF after the first maximum, the related error 
by the variance of the mean value. The formal error of the power law fit to the 
SF is determined by the variance of the fit parameters $a$ and $\beta$.  
The variability time scale is then determined by the crossing point 
of the power law and the fitted saturation level, $\rho_{0}=a\cdot t^{\beta}$, and
is given by
\begin{eqnarray}
t_{SF}=\left(\frac{\rho_{0}}{a}\right)^{1/\beta}\,\,,
\label{t_sf}
\end{eqnarray}
with its formal error
\begin{eqnarray}
\Delta t_{SF}=\left(\frac{\rho_{0}\pm\Delta\rho_{0}
/\sqrt3}{a\pm\Delta a/\sqrt3}\right)^{\frac{1}{\beta\pm\Delta\beta/\sqrt3}}\,.
\label{t_error}
\end{eqnarray}
In case of no clear saturation of the SF during the whole observing period, 
only a lower limit to the variability time scale is given (Table
\ref{timescales}, columns 2 and 3).\\

{\noindent \it (ii) Auto-correlation function analysis}\\
\\
In the investigation of pulsar scintillation, the usual way is to
define the ``de-correlation'' time scale as the time lag during which
the auto-correlation function (ACF) falls to 1/\emph{e} of its maximum 
\citep[e.g.][]{1986ApJ...311..183C}. The auto-correlation function 
$\rho(\tau)$ of a time series $S(t_{i})$ is defined as 
\citep[][]{1988ApJ...333..646E}
\begin{eqnarray}
\rho(\tau)=<(S(t)\cdot S(t-\tau))>_{t}\,\,.
\label{acf}
\end{eqnarray}
The ACF was calculated and $t_{ACF}$ was determined by the full-width 
of the ACF at 1/\emph{e}-height of its maximum. This method was applied to our 
data sets in total intensity and polarization. 
The error estimations were obtained taking the individual uncertainties of the calculated
auto-correlation functions into account. Columns 5 and 6 of Table \ref{timescales}
summarize the resulting characteristic time scales and error estimates.\\ 

{\noindent \it (iii) Minimum-maximum method}\\
\\
The observed total intensity variability patterns are mainly characterized 
by a quasi-monotonic increase of the flux density over a significant range of 
the total observing period. Consequently the SF/ACF methods deliver variability 
time scale of $3-4$ days, which owing to the limited total observing time must 
formally be regarded as lower limits to the true variability time scale $t_{var}$. 
In these cases we make use of the following definition of the time scale $t_{var}$ 
to obtain better estimates of the ``true'' variability time scales 
directly from those time series where only a several day trend was observed 
\citep[see also][and references therein]{1979ApJ...233..498M}: 
\begin{eqnarray}
t_{var}= \arrowvert\frac{d ln\,S}{dt}\,\arrowvert^{-1}=\frac{<S>\,\Delta t}{\Delta S}\,\,,
\label{t_min-max}
\end{eqnarray}
where $<S>$ denotes the mean flux density [Jy] and $\Delta S=S_{max}-S_{min}$ the flux 
density variation in the considered time interval $\Delta t$. 

We note that this time scale relies on a `visual' inspection of
the variability curves, defining by human eye (or mathematically 
via differentiation) the minimum and maximum of the flux density and 
the corresponding time interval $\Delta t$ between these two flux density 
levels. In Table \ref{timescales}, columns 8 \& 9 summarize the derived 
values for $\Delta t$ and $t_{var}$ of Eq. (\ref{t_min-max}).

\end{document}